\documentclass{siamart171218}
\usepackage{amsmath, amsfonts, amssymb}
\usepackage{xcolor}
\usepackage{graphicx}
\usepackage{subfig}
\usepackage{physics}

\graphicspath{{pdfimages/}}

\newcommand\iu{\mathrm{i}}
\newcommand{\Id}{\mathrm{Id}}

\newcommand{\EE}{\mathbb{E}}
\newcommand{\PP}{\mathbb{P}}

\newcommand{\CC}{\mathbb{C}}

\newcommand{\wt}[1]{\widetilde{#1}}

\newcommand{\mc}[1]{\mathcal{#1}}

\newsiamthm{remark}{Remark}

\title{A quantum kinetic Monte Carlo method for quantum many-body spin
  dynamics\thanks{This work is partially supported by the National Science
  Foundation under Grant Nos. DMS-1454939 and RNMS11-07444 (KI-Net).  Zhenning
  Cai is also supported by National University of Singapore Startup Fund under
  Grant No. R-146-000-241-133. The authors would like to thank Gero Friesecke
  for suggesting the problem and for illuminating discussions.}}

\author{Zhenning Cai\thanks{Department of Mathematics, National University of
  Singapore, Level 4, Block S17, 10 Lower Kent Ridge Road, Singapore 119076
  (\email{matcz@nus.edu.sg}).}
  \and Jianfeng Lu\thanks{Department of Mathematics, Department of Physics,
  Department of Chemistry, Duke University, Box 90320, Durham NC 27708, USA
  (\email{jianfeng@math.duke.edu}).}}

\begin{document}

\maketitle

\begin{abstract}
  We propose a general framework of quantum kinetic Monte Carlo
  algorithm, based on a stochastic representation of a series
  expansion of the quantum evolution. Two approaches have been
  developed in the context of quantum many-body spin dynamics, using
  different decomposition of the Hamiltonian. The effectiveness of the
  methods is tested for many-body spin systems up to $40$ spins.
\end{abstract}

\begin{keywords}
Spin dynamics, quantum kinetic Monte Carlo, Dyson series
\end{keywords}

\begin{AMS}
74S60
\end{AMS}

\section{Introduction}

We consider the system with $N$ spins in a magnetic field, which is
given by the Hamiltonian
\begin{equation}\label{eq:hamiltonian}
H(t) = \sum_{k=1}^N m^{(k)}(t) \cdot \sigma^{(k)} + \sum_{1 \leq j < k \leq N}
  \gamma_{jk}(t) \sigma^{(j)} \cdot \sigma^{(k)}
\end{equation}
with Hilbert space of the quantum system given by
$\mathcal{H} = (\mathbb{C}^2)^{\otimes N}$. For every
$k = 1,\cdots,N$, the operator $\sigma^{(k)}$ denotes the Pauli
matrices acting on the $k$-th spin:
\begin{equation}
\sigma^{(k)} = (\sigma_1^{(k)}, \sigma_2^{(k)}, \sigma_3^{(k)})^{\top}, \qquad
\sigma_i^{(k)} = \Id^{\otimes (k-1)} \otimes \sigma_i \otimes \Id^{\otimes (N-k)}; 
\end{equation}
where $\Id$ stands for the identity operator acting on a single spin
($\mathbb{C}^2$). The first term on the right hand side of
\eqref{eq:hamiltonian} gives the single-body Hamiltonians, where
$m^{(k)}(t) = (m_1^{(k)}(t), m_2^{(k)}(t), m_3^{(k)}(t))^{\top}$ is
the magnetic field acting on the $k$th spin. For the two-body
interaction in the Hamiltonian \eqref{eq:hamiltonian}, we have used
the notations
\begin{align}
  & \sigma^{(j)} \cdot \sigma^{(k)} = \sigma_1^{(j)} \sigma_1^{(k)} 
    + \sigma_2^{(j)} \sigma_2^{(k)} + \sigma_3^{(j)} \sigma_3^{(k)}, \\
  & \sigma_i^{(j)} \sigma_i^{(k)} = \Id^{\otimes (j-1)} \otimes \sigma_i \otimes \Id^{\otimes (k-j-1)} \otimes \sigma_i \otimes \Id^{\otimes (N-k)}.
\end{align}
Thus $\sigma^{(j)} \cdot \sigma^{(k)}$ counts for a Heisenberg type
interaction between the $j$-th and $k$-th spins with $\gamma_{jk}(t)$
being the interaction strength or coupling intensity. 

This paper concerns numerical algorithm for the time evolution of the
system: The many-body wave function $\ket{\Psi} \in \mathcal{H}$ is
govern by the Schr\"odinger equation
\begin{equation} \label{eq:Psi_eq} \frac{\mathrm{d}}{\mathrm{d} t}
  \ket{\Psi} = -\iu H \ket{\Psi}.
\end{equation}
While \eqref{eq:Psi_eq} is a linear ODE system, solving the system
directly is impractical even for dozens of spins, as the size of the
system $\dim \mathcal{H} = 2^N$ grows exponentially as the number of
spins increases. In fact, even representing a particular state
$\ket{\Psi}$ is challenging: for $N = 40$, the size of the vector is
greater than $1$ trillion; not mentioning the computational cost
involved in evaluation the matrix-vector product $H\ket{\Psi}$.

While the Hamiltonian \eqref{eq:hamiltonian} is rather general, our
algorithm development is mainly motivated by applications in nuclear
magnetic resonance (NMR) \cite{Ernst:90}, where the nuclear spins
react to magnetic fields.  In such applications, for the coefficient
of the single body term $m(t) = (m_1(t), m_2(t), m_3^{(k)})^{\top}$,
$(m_1(t), m_2(t))$ is prescribed as a control field (such that the
control magnetic field is only in $(x,y)$ direction) and the time
independent $m_3^{(k)}$ is understood as an energy splitting of the
$\ket{\uparrow}$ and $\ket{\downarrow}$ states of the $k$-th spin.
The $\gamma_{jk}$ terms account for dipole-dipole interactions between
the nuclear spins, and the magnitude of which decays very fast as the
distance between nuclei increases and is usually quite small compared
to the energy splitting and the external fields. The value of these
coefficients might contain some uncertainty due to experimental
imperfectness. One potential application of our method is robust
control of NMR via pulse design (see e.g., \cite{FrieseckeHennekeKunisch}
and references therein) when the spin-spin interactions are taken into
account, which we will leave for future works.

The high dimensionality of the system naturally calls for Monte Carlo
type methods. The motivation of the algorithm proposed in this work
comes from a surface hopping method recently developed by us in
\cite{CaiLu2017}, which can be viewed as a stochastic method to solve
generic high dimensional ODE systems (or PDE systems combined with
some particle / semiclassical methods, as in \cite{LuZhou2016,
  CaiLu2017}).  The overall idea of the algorithm contains two
elements:
\begin{enumerate}
\item a series expansion of the solution of the system based on a
  time-dependent perturbation theory, \textit{i.e.}, from the ODE
  point of view, a repeated back-substitution in the integral form of
  the system based on Duhamel's principle;
\item a Monte Carlo method to stochastically evaluate the series
  expansion based on an efficient representation of part of the
  Hilbert space.
\end{enumerate}
In this work, we will apply the above framework to develop methods for
quantum many-body spin dynamics. For a particular system, to make the
algorithm efficient, it is crucial to identify a suitable ``small
term'' to be used in the series expansion from time-dependent
perturbation theory. In the setting of spin dynamics, we will discuss
two approaches: 1) one is based on a decomposition of the Hamiltonian
into terms commuting with $\sigma_3$ and those not commuting
(\textit{i.e.}, diagonal and off-diagonal terms in the $Z$-basis of
the spins); 2) the other approach is based on a splitting of the
Hamiltonian into single-body term and two-body interactions.

Over the years, many numerical methods have been developed in physics
and chemistry literature for many-body quantum dynamics, which is a
central challenge in theoretical understanding of quantum
systems. While a complete literature review is beyond the scope, we
discuss here some related works to our approach. 

The methods proposed in the literature can be roughly categorized into
two groups.  One class of methods is based on an efficient
representation of the relevant part of the Hilbert space of the
many-body quantum system, such as the multi-configurational
time-dependent Hartree (MCTDH) \cite{MeyerMantheCederbaum:1990},
originally developed for the nucleus dynamics, which uses
multi-configurational Hartree ansatz to represent wave functions.
Other methods belong to this class include the time-evolving block
decimation (TEBD) \cite{Vidal:2003} and the time-dependent density
matrix renormalization group (tDMRG) \cite{Schollwock:2011} as
extensions of the DMRG method \cite{White:1992} to dynamical
problems. These methods are based on matrix product states and hence
rather powerful for one (physical) dimensional systems, but face
difficulty in extending to higher dimensions.

The other class of methods is based on Monte Carlo sampling. A
particular relevant class of quantum Monte Carlo methods to our method
is the continuous-time quantum Monte Carlo (CT-QMC)
\cite{RubtsovSavkinLichtenstein:2005, Prokofev:96, Gull:2011}, which
can be understood as a continuum time limit of the Trotter splitting
based Hirsch-Fye QMC method \cite{HirschFye:1986}. While the CT-QMC
method was originally developed for imaginary time propagation, it has
also been extended to real time dynamics in recent years
\cite{MuhlbacherRabani:2008, CohenGullReichmanMillis:2015}, in
particular for impurity models in condensed matter physics. Another
related quantum Monte Carlo method is the auxiliary field quantum
Monte Carlo \cite{ZhangKrakauer:2003, Zhang:2013} based on the
Hubbard-Stratonovich transformation \cite{Hubbard:1959,
  Stratonovich:1958}, which represents the imaginary time evolution of
the many-body electronic wave function as a stochastic sum of Slater
determinants. We also note a different strategy -- discrete truncated
Wigner approximation (DTWA) -- proposed recently
\cite{Schachenmayer:2015} based on a tensor product ansatz on the
level of discrete Wigner representation of many-body spin density
matrix. The Monte Carlo method is used to sample initial states
according to the phase space distribution.

Our method can be understood at the interface of the above two
categories: The idea of series expansion is also utilized in the
CT-QMC method. Unlike CT-QMC, which resorts to diagrammatic
perturbation ideas in many-body theory, our stochastic evaluation
method is based on ansatz representation similar to those used in
MCTDH. The proposed method is not restricted to a particular
  geometry of the systems, and is expected to work well when the
stochastic ansatz captures the behavior of the physical system under
study, as will be further illustrated by numerical examples. As
another goal of the manuscript, we hope that the abstraction of the
ideas developed in the physics and chemistry literature would help
transferring these techniques for high dimensional computational
challenges we face in other areas. Indeed, the unified framework of
our method can be applied to any linear evolution problems in high
dimensions.

\section{Algorithm}

\subsection{Quantum kinetic Monte Carlo algorithm}

Before we turn to the specific quantum many-body spin dynamics, let us
present the general framework of the algorithm. This is an abstraction
of the ideas behind the surface hopping algorithms developed in our
previous works~\cite{LuZhou2016, CaiLu2017}.

Given a Hamiltonian $H$ on the Hilbert space $\mc{H}$, the quantum
kinetic Monte Carlo algorithm starts with a choice of a decomposition
of the Hamiltonian
\begin{equation} \label{eq:decomposition}
  H = H_{\text{easy}} + H_{\text{hard}},
\end{equation}
together with a class of states $\mc{A} \subset \mc{H}$. We require
that
\begin{enumerate}
\item[A)] Any vector $\ket{\Psi} \in \mc{A}$ is easy to represent
  (i.e., we do not need to store directly the full vector, but only a
  parametrized form of it). In general, $\mc{A}$ might not be a vector
  space, i.e., the parametrization is nonlinear;
\item[B)] For any $\ket{\Psi} \in \mc{A}$, the action of
  $H_{\text{easy}}$ remains in $\mc{A}$:
  \begin{equation}
    \mathrm{e}^{-\iu t H_{\text{easy}}} \ket{\Psi} \in \mc{A}, \qquad \forall\, t 
  \end{equation}
  and is easy to obtain (either exactly or with a controllable error);
\item[C)] It is possible to stochastically represent the action of
  $H_{\text{hard}}$, in the sense that there exists a stochastic
  operator $A(\omega)$ with $\omega$ corresponding to some random
  space $\Omega$, such that for any $\ket{\Psi} \in \mc{A}$,
  \begin{equation} \label{eq:hard_op}
    \EE_{\omega} A(\omega) \ket{\Psi} = H_{\text{hard}} \ket{\Psi}, \qquad \text{and} \qquad A(\omega) \ket{\Psi} \in \mc{A}, \quad \forall \omega.
  \end{equation}
\end{enumerate}
In practice, it is often easier to first determine the set $\mc{A}$,
and then look for the decomposition of the Hamiltonian and also the
stochastic representation of the action of $H_{\text{hard}}$. Thus, in
the following, when discussing two examples of algorithms developed
under this framework, we will refer to them by the choice of the set
$\mc{A}$.

With such a decomposition, the Schr\"odinger equation
\eqref{eq:Psi_eq} can be written as the following integral form by the
Duhamel's principle:
\begin{equation} \label{eq:int_form}
\begin{split}
\ket{\Psi(t)} &= \mathrm{e}^{-\iu t H_{\mathrm{easy}}} \ket{\Psi(0)} -
  \iu \int_0^t \mathrm{e}^{-\iu (t-t_1) H_{\mathrm{easy}}}
    H_{\mathrm{hard}} \ket{\Psi(t_1)} \,\mathrm{d}t_1 \\
&= \mathrm{e}^{-\iu t H_{\mathrm{easy}}} \ket{\Psi(0)} -
  \iu \int_0^t \int_{\Omega} \mathrm{e}^{-\iu (t-t_1) H_{\mathrm{easy}}}
    A(\omega_1) \ket{\Psi(t_1)} \,\mathrm{d}\mu_{\omega_1} \,\mathrm{d}t_1,
\end{split}
\end{equation}
where we have used $\mu_{\omega}$ to denote the probability measure of $\omega$ and write 
\begin{equation}
  \EE_{\omega} f(\omega) = \int_{\Omega} f(\omega) \,\mathrm{d} \mu_{\omega}.
\end{equation}
Note that the right hand side of \eqref{eq:int_form} involves the
unknown wave function at time $t_1$. To proceed, we apply
\eqref{eq:int_form} to its own right hand side and get
\begin{equation} \label{eq:int_form2}
\begin{split}
& \ket{\Psi(t)} = \mathrm{e}^{-\iu t H_{\mathrm{easy}}} \ket{\Psi(0)} -
  \iu \int_0^t \int_{\Omega} \mathrm{e}^{-\iu (t-t_1) H_{\mathrm{easy}}}
    A(\omega_1) \mathrm{e}^{-\iu t_1 H_{\mathrm{easy}}} \ket{\Psi(0)} 
  \,\mathrm{d}\mu_{\omega} \,\mathrm{d}t_1 \\
& + (-\iu)^2 \int_0^t \!\! \int_{\Omega} \int_0^{t_1} \!\!\! \int_{\Omega}
  \mathrm{e}^{-\iu (t-t_1) H_{\mathrm{easy}}} A(\omega_1)
  \mathrm{e}^{-\iu (t_1-t_2) H_{\mathrm{easy}}} A(\omega_2) \ket{\Psi(t_2)}
  \,\mathrm{d}\mu_{\omega_2} \,\mathrm{d}t_2 \,\mathrm{d}\mu_{\omega_1} \,\mathrm{d}t_1.
\end{split}
\end{equation}
Inserting again \eqref{eq:int_form} into the right hand side of
\eqref{eq:int_form2}, we will get one more term on the right hand
side. Such a substitution can be done repeatedly, which results into
the following Dyson series expansion of $\ket{\Psi}$:
\begin{equation} \label{eq:Psi_expansion}
\begin{split}
  \ket{\Psi(t)} &= \sum_{M=0}^{+\infty} \int_0^t
  \int_{\Omega} \int_0^{t_1} \int_{\Omega} \cdots
  \int_0^{t_{M-1}} \int_{\Omega} (-\iu)^M \times \\
  & \times \mathrm{e}^{-\iu (t-t_1) H_{\mathrm{easy}}}
  A(\omega_1) \mathrm{e}^{-\iu (t_1-t_2) H_{\mathrm{easy}}}
  A(\omega_2) \cdots \\
  & \times \mathrm{e}^{-\iu (t_{M-1}-t_M)
    H_{\mathrm{easy}}} A(\omega_M) \mathrm{e}^{-\iu t_M
    H_{\mathrm{easy}}} \ket{\Psi(0)} \,\mathrm{d}\mu_{\omega_M}
  \,\mathrm{d}t_M \cdots
  \,\mathrm{d}\mu_{\omega_2} \,\mathrm{d}t_2 \,\mathrm{d}\mu_{\omega_1} \,\mathrm{d}t_1 \\
  &= \sum_{M=0}^{+\infty} \int_0^t \int_{\Omega} \int_0^{t_M}
  \int_{\Omega} \cdots
  \int_0^{t_2} \int_{\Omega} (-\iu)^M \times \\
  & \times \mathrm{e}^{-\iu (t-t_M) H_{\mathrm{easy}}}
  A(\omega_M) \mathrm{e}^{-\iu (t_M-t_{M-1}) H_{\mathrm{easy}}}
  A(\omega_{M-1}) \cdots \\
  & \times \mathrm{e}^{-\iu (t_2-t_1) H_{\mathrm{easy}}}
  A(\omega_1) \mathrm{e}^{-\iu t_1 H_{\mathrm{easy}}} \ket{\Psi(0)}
  \,\mathrm{d}\mu_{\omega_1} \,\mathrm{d}t_1 \cdots \,\mathrm{d}\mu_{\omega_{M-1}}
  \,\mathrm{d}t_{M-1} \,\mathrm{d}\mu_{\omega_M} \,\mathrm{d}t_M,
\end{split}
\end{equation}
where the last step is just renaming the integration variables. Using
dominated convergence, it is easy to see that if the operators
$A(\omega)$ are uniformly bounded, the above series expansion
converges absolutely.  According to the properties B) and C), the
integrand in the above equation is always a state in $\mc{A}$, and is
therefore easy to represent. The challenge then lies in the high
dimensional integration both in the stochastic space and in the time
sequence, for which we turn to Monte Carlo method.

Before discussing the algorithm, let us note that the above can be
extended to time-dependent Hamiltonian operators, with the assumption
\eqref{eq:hard_op} now changes to for any $\ket{\Psi} \in \mc{A}$
\begin{equation} 
  \EE_{\omega} A(t,\omega) \ket{\Psi} = H_{\mathrm{hard}}(t) \ket{\Psi}, \qquad \text{and}  \qquad A(t, \omega) \ket{\Psi} \in \mc{A},
\end{equation}
and the semigroup generated by $H_{\mathrm{easy}}(t)$ preserves the
state in $\mc{A}$ (it is also possible to consider the more general
case that the set $\mc{A}$ depends on time).  The expansion
\eqref{eq:Psi_expansion} changes to
\begin{equation} \label{eq:Psi_expansion_td}
\begin{split}
  \ket{\Psi(t)} &= \sum_{M=0}^{+\infty} \int_0^t
  \int_{\Omega} \int_0^{t_M} \int_{\Omega} \cdots
  \int_0^{t_2} \int_{\Omega} (-\iu)^M \times \\
  & \qquad \times U(t,t_M) A(t_M,\omega_M)
  U(t_M,t_{M-1}) A(t_{M-1},\omega_{M-1}) \cdots \\
  & \qquad \times U(t_2,t_1) A(t_1,\omega_1) U(t_1,0)
  \ket{\Psi(0)} \,\mathrm{d}\mu_{\omega_1} \,\mathrm{d}t_1 \cdots
  \,\mathrm{d}\mu_{\omega_{M-1}} \,\mathrm{d}t_{M-1} \,\mathrm{d}\mu_{\omega_M}
  \,\mathrm{d}t_M.
\end{split}
\end{equation}
with the unitary evolution operator $U(t, s)$ given by
\begin{equation} \label{eq:U_definition}
U(t,s) := \mc{T}
  \exp \left( -\iu \int_s^t H_{\mathrm{easy}}(\tau) \,\mathrm{d}\tau \right), 
\end{equation}
where $\mc{T}$ is the time-ordering operator.

\medskip

The expansion \eqref{eq:Psi_expansion_td} inspires us to use Monte Carlo method
to evaluate $\ket{\Psi(t)}$. In such a method, each sample would be the
integrand on the right hand side of \eqref{eq:Psi_expansion_td}. To determine
the integrand, we need to specify the following:
\begin{enumerate}
\item A non-negative integer $M$;
\item A sequence of random times:
  $0 \leq t_1 \leq t_2 \leq \cdots \leq t_M \leq t$;
\item A sequence of random samples in $\Omega$: $\omega_1, \cdots, \omega_M$.
\end{enumerate}
This links the Monte Carlo method to a marked point
process with mark space $\Omega$ (see the textbook \cite{Jacobsen2006}
for an introduction of the marked point process and also point process
in general). We denote $\Xi = ((t_m), (\omega_m))_{m\geqslant 1}$ one
realization of the marked point process, in which
$\omega_m \in \Omega$ is marked at time $t_m$. The marked point
process is generated by an intensity function $\lambda(t,\omega)$, \textit{i.e.}
\begin{equation}
\mathbb{P}(\text{A mark in } \Sigma \text{ appears in } [t,t+h)) =
  \int_{\Sigma} \lambda(t,\omega) h \,\mathrm{d}\mu_{\omega} + o(h),
  \qquad \forall t > 0, \quad \forall \Sigma \subset \Omega.
\end{equation}
Any intensity function that is strictly positive can be used, a better
choice will however reduce the sampling variance of the algorithm. We will further discuss the choice of the intensity function in the next section.

The following identity is essential to our method: It turns the
problem of calculating $\ket{\Psi(t)}$ into a sampling problem:
\begin{equation} \label{eq:expectation}
\ket{\Psi(t)} = \EE_{\Xi} \ket{\Phi_{\Xi}(t)},
\end{equation}
if for a given realization of the marked point process $\Xi$, the state
$\ket{\Phi_{\Xi}(t)}$ is defined by
\begin{equation} \label{eq:traj_def}
\begin{split}
\ket{\Phi_{\Xi}(t)} = & \exp \left(
  \int_0^t\int_{\Omega} \lambda(s,\omega) \,\mathrm{d}\mu_{\omega} \,\mathrm{d}s
\right) U(t,t_M) \wt{A}(t_M,\omega_M) \times \\
& \qquad \times U(t_M, t_{M-1}) \wt{A}(t_{M-1}, \omega_{M-1}) \cdots
  U(t_2,t_1) \wt{A}(t_1,\omega_1) U(t_1,0) \ket{\Psi(0)},
\end{split}
\end{equation}
where $M$ is the number of marks in $\Xi$ before time $t$, and we have
used the short hands
\begin{equation} \label{eq:hat_A}
  \wt{A}(t_m, \omega_m) = -\iu A(t_m, \omega_m) / \lambda(t_m,\omega_m),
  \qquad m = 1, \cdots, M.
\end{equation}

The equality \eqref{eq:expectation} follows as for any function
$\mc{F}(\Xi)$ depending only on the part of $\Xi$ in the time interval
$[0,t)$, we have
\begin{equation} \label{eq:expectation_F}
\begin{split}
\EE_{\Xi} \mc{F}(\Xi) &= \sum_{M=0}^{+\infty} \int_0^t \int_{\Omega}
  \int_0^{t_M} \int_{\Omega} \cdots \int_0^{t_2} \int_{\Omega}
\exp \left(
  -\int_0^t\int_{\Omega} \lambda(s,\omega) \,\mathrm{d}\mu_{\omega} \,\mathrm{d}s
\right) \times \\
& \qquad \times \left( \prod_{m=1}^M \lambda(t_m, \omega_m) \right) \mc{F}(\Xi)
  \,\mathrm{d}\mu_{\omega_1} \,\mathrm{d}t_1 \cdots
  \,\mathrm{d}\mu_{\omega_{M-1}} \,\mathrm{d}t_{M-1} \,\mathrm{d}\mu_{\omega_M}
  \,\mathrm{d}t_M.
\end{split}
\end{equation}


\newcommand\tPhi{\lvert \wt{\Phi}_{\Xi}(t)\rangle}
An algorithm to evaluate $\ket{\Psi(t)}$ naturally follows the
equality \eqref{eq:expectation}. The idea is to draw a sequence of
realizations of process $\Xi$, evaluate $\ket{\Phi_{\Xi}(t)}$ for each
realization, an estimate of $\ket{\Psi}$ is then given by the average. For
easier implementation, we define
\begin{equation} \label{eq:eta_def}
\eta(t) = \int_0^t\int_{\Omega}
  \lambda(s,\omega) \,\mathrm{d}\mu_{\omega} \,\mathrm{d}s, \qquad
\tPhi = \mathrm{e}^{-\eta(t)} \ket{\Phi_{\Xi}(t)}.
\end{equation}
It can be easily seen that $\eta(t)$ can be obtained by solving
\begin{equation} \label{eq:eta}
\frac{\mathrm{d}\eta}{\mathrm{d}t} = \int_{\Omega}
  \lambda(t,\omega) \,\mathrm{d}\mu_{\omega}.
\end{equation}
Thus, while evolving the state $\ket{\Phi_{\Xi}(t)}$, an additional
scalar quantity $\eta(t)$ needs to be evolved simultaneously. In our
application, the mark space $\Omega$ is a finite set, and therefore solving
\eqref{eq:eta} does not introduce much numerical cost. In \eqref{eq:eta_def},
the time-dependent state $\tPhi$ can be considered as a trajectory in $\mc{A}$.
To obtain this trajectory, we use the fact that between two adjacent marks, the
trajectory satisfies
\begin{equation} \label{eq:tilde_Phi}
\frac{\mathrm{d}}{\mathrm{d}t} \tPhi = -\iu H_{\mathrm{easy}}(t) \tPhi,
  \qquad t \in (t_{m-1}, t_m), \quad m > 0.
\end{equation}
When a mark is met, one just needs to apply the operator $\wt{A}(t_m,
\omega_m)$. An illustration of the trajectory $\tPhi$ is given in Figure
\ref{fig:tPhi}.
\begin{figure}[!ht]
\centering
\includegraphics[width=.7\textwidth]{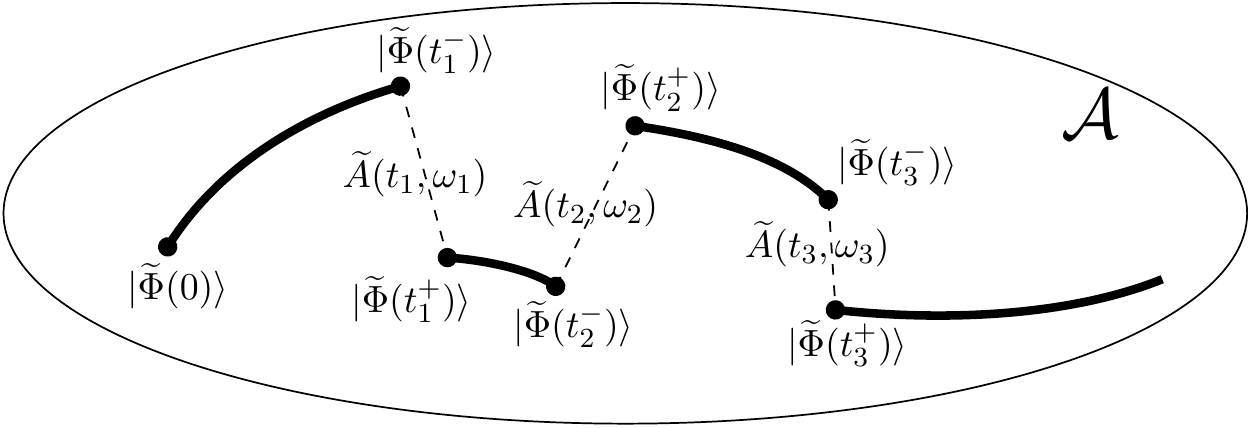}
\caption{Illustration of the trajectory $\tPhi$. The bold lines denote the
evolution of the equation \eqref{eq:tilde_Phi}, and the thin dashed lines
denote the application of the operator marked over them. The states $\lvert
\wt{\Phi}_{\Xi}(t_k^{\pm})\rangle$ are the left/right limits of $\tPhi$ at
$t_k$.}
\label{fig:tPhi}
\end{figure}

In \eqref{eq:eta_def}, the reason for introducing $\eta(t)$ is the following
property:
\begin{equation} \label{eq:prob_no_mark}
\PP(\text{no mark exists in } (s_1, s_2)) =
  \exp \big( -[\eta(s_2) - \eta(s_1)] \big),
\end{equation}
which helps us to realize the marked point process. In our implementation, the
drawing of the marked point process and the evolution of the trajectory are
simultaneously done. In detail, one trajectory $\tPhi$ can be obtained by the
following steps:
\begin{enumerate}
\item Set $t \leftarrow 0$, $\tilde{\eta} \leftarrow 0$,
  $|\wt{\Phi}_{\Xi}(0) \rangle \leftarrow \ket{\Psi(0)}$, $\mathit{flag}
  \leftarrow \mathit{false}$. Generate a random number $Y$ obeying the uniform
  distribution in $[0,1]$.
\item Stop if $t$ is large enough. Otherwise, select a time step $\Delta t$,
  and solve $\eta(t + \Delta t)$ according to \eqref{eq:eta}.
\item If $\exp(\tilde{\eta} - \eta(t + \Delta t)) \leqslant 1 - Y$, then
  solve the equation of the mark time $\tilde{t̃}$
  \begin{displaymath}
  \exp(\tilde{\eta} - \eta(\tilde{t})) = 1 - Y
  \end{displaymath}
  by interpolation of $\eta(t)$ in $[t, t + \Delta t]$, and set $\Delta t
  \leftarrow \tilde{t} - t$, $\mathit{flag} \leftarrow \mathit{true}$,
  $\tilde{\eta} = \eta(\tilde{t})$.
\item Solve the equation \eqref{eq:tilde_Phi} to get $|\wt{\Phi}_{\Xi}
  (t+\Delta t) \rangle$. Set $t \leftarrow t + \Delta t$.
\item If $\mathit{flag}$ is $\mathit{false}$, return to step 2. Otherwise,
  generate a mark $\omega \in \Omega$ according to the probability measure
  $\mu_{\omega}$, and set $\tPhi \leftarrow \wt{A}(t,\omega) \tPhi$. Generate
  a new uniformly distributed random variable $Y \in [0,1]$ and return to step
  2.
\end{enumerate}
In the above algorithm, $\tilde{\eta}$ records the value of $\eta$ at the
last mark, and step 3 uses the property \eqref{eq:prob_no_mark} to determine
the time of the next mark. When a mark is set at the current time step, the
boolean variable $\mathit{flag}$ is set to be true, and in step 5, a mark is
picked from the mark space. By doing this, the whole process $\Xi$ is generated
mark by mark. At the same time, the value of $\eta(t)$ at the trajectory
$\tPhi$ is obtained at all discrete times, and thus $\ket{\Phi_{\Xi}(t)}$ can
be evaluated by $\ket{\Phi_{\Xi}(t)} = \mathrm{e}^{\eta(t)} \tPhi$.
Regarding the computational cost, it can be seen that at each time step
for a single trajectory, we need to evaluate the integral over $\Omega$ for
$O(1)$ times (depends on the Runge-Kutta method used), apply $H_{\mathrm{easy}}$
for $O(1)$ times, and apply $\wt{A}(t,\omega)$ at most once.

\begin{remark}
  This section gives a general framework for the algorithm. To apply
  the algorithm, we need a decomposition \eqref{eq:decomposition}
  satisfying the conditions A) to C). In general, such decomposition
  should be studied case by case. One standard way is to decompose the
  Hamiltonian into non-interacting and interacting parts, where the
  non-interacting part will be regarded as $H_{\mathrm{easy}}$ and the
  interacting part as $H_{\mathrm{hard}}$. Since the interacting part
  usually induces many-body entanglement to the system and is
  therefore considered as ``hard''.
\end{remark}

\subsection{QKMC with $Z$-basis} \label{sec:Z-basis}

Let us now consider specific examples of the quantum kinetic Monte Carlo algorithm. In the first example, we take 
\begin{equation}
  \mc{A} = \Bigl\{\ket{\Psi} =  a \ket{\psi_1} \otimes \cdots \otimes \ket{\psi_N} \;\big\vert\; a \in \CC,\, \ket{\psi_i} \in \{ \ket{\uparrow}, \ket{\downarrow} \}, \, \forall i
  \Bigr\}
\end{equation}
We call this the $Z$-basis, since each $\ket{\Psi} \in \mc{A}$ is an
eigenvector of (one-body or many-body) Pauli matrices in the
$z$-direction.

It is obvious that we can choose $H_{\mathrm{easy}}$ as
\begin{equation}
H_{\mathrm{easy}}(t) = \sum_{k=1}^N m_3^{(k)} \sigma_3^{(k)}
  + \sum_{1 \le j < k \le N} \gamma_{jk}(t) \sigma_3^{(j)} \sigma_3^{(k)},
\end{equation}
since for $U(t,s)$ defined as \eqref{eq:U_definition},
\begin{equation}
  U(t,s) \ket{\Psi} = \exp\left(
    -\iu (t-s) \sum_{k=1}^N \kappa_k m_3^{(k)}
    -\iu \sum_{1\le j<k \le N} \kappa_j \kappa_k
      \int_s^t \gamma_{jk}(\tau) \,\mathrm{d}\tau
  \right) \ket{\Psi},
\end{equation}
where $\ket{\Psi}$ is an arbitrary state in $\mc{A}$ and $\kappa_k =
\bra{\psi_k} \sigma_3 \ket{\psi_k}$. The sample space for $\omega$ can be
chosen as $\Omega = \Omega_1 \cup \Omega_2$, where
\begin{equation}
\Omega_1 = \{ k \mid k = 1,\cdots,N \}, \quad
\Omega_2 = \{ (j,k) \mid j,k = 1,\cdots,N, \, j < k \},
\end{equation}
and the operator $A(t, \omega)$ is set to be
\begin{equation}
A(t, \omega) = \left\{ \begin{array}{ll}
  |\Omega| (m_1^{(k)} \sigma_1^{(k)} + m_2^{(k)} \sigma_2^{(k)}),
    & \text{if } \omega = k \in \Omega_1, \\
  |\Omega| \gamma_{jk}(t) (\sigma_1^{(j)} \sigma_1^{(k)} + \sigma_2^{(j)} \sigma_2^{(k)}),
    & \text{if } \omega = (j,k) \in \Omega_2,
\end{array} \right.
\end{equation}
where $|\Omega| = N(N+1)/2$ is the cardinality of $\Omega$. For all
$\omega_0 \in \Omega$, the probabilities are assigned the same:
\begin{equation}
\PP(\omega = \omega_0) = \frac{1}{\abs{\Omega}},
\end{equation}
which gives the defintion of $\mu_{\omega}$. The above definitions immediately
lead to
\begin{equation}
\EE_{\omega} A(t,\omega) = H_{\mathrm{hard}} := H - H_{\mathrm{easy}}.
\end{equation}
In our implementation, the intensity function $\lambda(t,\omega)$ is chosen as
\begin{equation}
\lambda(t, \omega) = \left\{ \begin{array}{ll}
  |\Omega| \bigl\lvert m_1^{(k)} + \iu m_2^{(k)} \bigr\rvert
    & \text{if } \omega = k \in \Omega_1, \\[5pt]
  2 |\Omega| |\gamma_{jk}(t)| & \text{if } \omega = (j,k) \in \Omega_2,
\end{array} \right.
\end{equation}
so that the operator $\wt{A}(t,\omega)$ defined in \eqref{eq:hat_A} does not
change the magnitude of the state.

\begin{remark}
The $Z$-basis can be considered as a weight multiplied by an element in the
finite set
\begin{equation}
\mc{S} = \left\{ \ket{\psi_1} \otimes \cdots \otimes \ket{\psi_N}
  \Big\vert \ket{\psi_i} \in \{ \ket{\uparrow}, \ket{\downarrow} \}, \forall i
\right\}.
\end{equation}
From this point of view, the marked point process can be interpreted
as a jump process with state space $\mc{S}$. In detail, supposing the
current state in the jump process is $\ket{\Psi}$, we can interpret
the mark $\omega$ at time $t$ as the jump
$\ket{\Psi} \rightarrow \ket{\Phi}$, where $\ket{\Phi}$ is the only
state in $\mc{S}$ such that
$\bra{\Phi} A(t,\omega) \ket{\Psi} \neq 0$. At each jump, the weight
is changed to fit the result of applying  $\wt{A}(t, \omega)$.
Such an interpretation matches the currently proposed method with the
surface hopping method proposed in \cite{CaiLu2017}, where each
element in $\mc{S}$ is considered as a ``surface'', and the weight is
considered as evolving on the surfaces.
\end{remark}

\subsection{QKMC with simple tensors}

In this case, we choose the subset $\mc{A}$ to be all simple tensors in the tensor product space
$\mc{H} = (\CC^2)^{\otimes N}$.
\begin{equation}
  \mc{A} = \Bigl\{\ket{\Psi} =  \ket{\psi_1} \otimes \cdots \otimes \ket{\psi_N} \;\big\vert\; \ket{\psi_i} \in \CC^2, \, \forall i
  \Bigr\}
\end{equation}
It is clear that this set is larger than the one of $Z$-basis.

In such a case, we can choose $H_{\mathrm{easy}}$ to be
\begin{equation}
H_{\mathrm{easy}}(t) = \sum_{k=1}^N m^{(k)}(t) \cdot \sigma^{(k)}.
\end{equation}
Thus for any $\ket{\Psi} = \ket{\psi_1} \otimes \cdots \otimes \ket{\psi_N} \in
\mc{A}$, we have
\begin{displaymath}
U(t,s) \ket{\Psi} =
  U^{(1)}(t,s) \ket{\psi_1} \otimes \cdots \otimes U^{(N)}(t,s) \ket{\psi_N}
  \in \mc{A},
\end{displaymath}
where
\begin{displaymath}
U^{(k)}(t,s) = \exp \left(
  -\iu \left[ \int_s^t m^{(k)}(\tau) \,\mathrm{d}\tau \right] \cdot \sigma
\right),
\end{displaymath}
where $\sigma = (\sigma_1, \sigma_2, \sigma_3)^{\top}$. The sample space
$\Omega$ and the probability measure $\mu_{\omega}$ are given by
\begin{equation} \label{eq:Omega}
\Omega = \{(i,j,k) \mid i = 1,2,3, \, j,k = 1,\cdots,N, \, j < k \}, \qquad
\PP(\omega = \omega_0) = 1 / |\Omega|, \quad \forall \omega_0 \in \Omega.
\end{equation}
By defining
\begin{equation} \label{eq:A}
A(t,\omega) = |\Omega| \gamma_{jk}(t) \sigma_i^{(j)} \sigma_i^{(k)},
\end{equation}
it is easy to find that $\EE_{\omega} A(t,\omega) = H(t) -
H_{\mathrm{easy}}(t)$. Again, we choose
\begin{equation} \label{eq:lambda}
\lambda(t,\omega) = |\Omega| |\gamma_{jk}(t)|,
\end{equation}
so that $\wt{A}(t,\omega)$ does not change the norm of the state.

In this method, the evolution equation for $\eta(t)$ defined in
\eqref{eq:eta_def} is
\begin{equation} \label{eq:eta1}
\frac{\mathrm{d}\eta}{\mathrm{d}t} =
  3 \sum_{1\le j < k \le N} |\gamma_{jk}(t)|.
\end{equation}
In the method using $Z$-basis (see Section \ref{sec:Z-basis}), we have
\begin{equation} \label{eq:eta2}
\frac{\mathrm{d}\eta}{\mathrm{d}t} =
  \sum_{k=1}^N \bigl\lvert m_1^{(k)} + \iu m_2^{(k)} \bigr\rvert
  + 2 \sum_{1\le j < k \le N} |\gamma_{jk}(t)|.
\end{equation}
Therefore, if the coupling intensity is significantly smaller than the control
field, the growing rate of $\eta(t)$ in \eqref{eq:eta2} is larger than that in
\eqref{eq:eta1}, which indicates larger variance in the numerical solution with
$Z$-basis.

\section{Analysis of the sampling variance}
In this section, we discuss the evolution of the numerical error for the above
algorithm. Suppose $N_{\mathrm{traj}}$ trajectories are used in the simulation.
Then the numerical solution is
\begin{equation} \label{eq:Psi_num}
\ket{\Psi_{\mathrm{num}}(t)} = \frac{1}{N_{\mathrm{traj}}}
  \sum_{i=1}^{N_{\mathrm{traj}}} \ket{\Phi_{\Xi_i}(t)},
\end{equation}
where $\Xi_i$ is the realization of the marked point process corresponding to
the $i$-th trajectory, and $\Xi_i$ and $\Xi_j$ are independent of each other if
$i \neq j$. Define $\ket{\Psi_{\mathrm{err}}(t)}$ as the difference between the
numerical solution and the exact solution:
\begin{equation} \label{eq:Psi_err}
\ket{\Psi^{\mathrm{err}}(t)} =
  \ket{\Psi^{\mathrm{num}}(t)} - \ket{\Psi(t)}.
\end{equation}
Apparently $\EE \ket{\Psi^{\mathrm{err}}(t)} = 0$. Therefore the variance of
$\ket{\Psi^{\mathrm{err}}(t)}$ is
\begin{equation} \label{eq:sigma}
\sigma^2 = \EE \braket{\Psi^{\mathrm{err}}(t)},
\end{equation}
which will be estimated below.

Inserting \eqref{eq:Psi_num} and \eqref{eq:Psi_err} into \eqref{eq:sigma} and
using the fact that $\EE_{\Xi_i} \ket{\Phi_{\Xi_i}(t)} = \ket{\Psi(t)}$ for any
$i$, we have
\begin{equation}
\sigma^2 = \frac{1}{N_{\mathrm{traj}}^2}
  \sum_{i=1}^{N_{\mathrm{traj}}} \sum_{j=1}^{N_{\mathrm{traj}}}
    \EE_{\Xi_i,\Xi_j} \braket{\Phi_{\Xi_i}(t)}{\Phi_{\Xi_j}(t)} - \braket{\Psi(t)}.
\end{equation}
Since all trajectories are independent, in the above sum, the terms with $i
\neq j$ can be directly evaluated, which yields
\begin{equation}
\sigma^2 = \frac{1}{N_{\mathrm{traj}}} \EE_{\Xi} \braket{\Phi_{\Xi}(t)}
  - \frac{1}{N_{\mathrm{traj}}} \braket{\Psi(t)}.
\end{equation}
This equation shows that the the variance is proportional to the
inverse of the number of trajectories, which is of course typical for
Monte Carlo algorithms.

To further estimate $\EE_{\Xi} \braket{\Phi_{\Xi}(t)}$, we use the definition
of the trajectory \eqref{eq:traj_def} to get
\begin{equation} \label{eq:Phi_norm}
\braket{\Phi_{\Xi}(t)} \leqslant
  \exp \left( 2\int_0^t \int_{\Omega} \lambda(s,\omega)
    \,\mathrm{d}\mu_{\omega} \,\mathrm{d}s \right)
  \left( \prod_{m=1}^M \| \wt{A}(t_m, \omega_m) \| \right)^2
  \braket{\Psi(0)},
\end{equation}
where $\| \cdot \|$ is the operator norm, and we have used the fact that
$U(\cdot, \cdot)$ is a unitary operator. If the intensity function
$\lambda(t,\omega)$ is chosen such that there exist constants $\Lambda$ and
$\wt{\alpha}$ satisfying
\begin{equation} \label{eq:Lam_al}
|\lambda(t,\omega)| \leqslant \Lambda \quad \text{and}
  \quad \|\wt{A}(t,\omega)\| \leqslant \wt{\alpha}, \qquad
  \forall t \in \mathbb{R}^+, \quad \forall \omega \in \Omega,
\end{equation}
we can apply \eqref{eq:expectation_F} to get
\begin{equation} \label{eq:estimation}
\EE_{\Xi} \braket{\Phi_{\Xi}(t)} \leqslant
  \mathrm{e}^{\Lambda t} \braket{\Psi(0)} \sum_{M=0}^{+\infty}
    \frac{t^M}{M!} \Lambda^M \wt{\alpha}^{2M}
  = \mathrm{e}^{\Lambda(1+\wt{\alpha}^2)t} \braket{\Psi(0)}.
\end{equation}
Here one sees that the variance grows exponentially with respect to $t$.

Note that the equation \eqref{eq:Phi_norm} becomes an equality if every
operator $A(t,\omega)$ is a unitary operator multiplied by a positive constant
$\alpha$.  This condition holds for \eqref{eq:A} if we have either
$\gamma_{jk}(t) \equiv 0$ or $\gamma_{jk}(t) \equiv \gamma$ for any $j,k$, and
in this case, we have $\alpha = \norm{A(t, \omega)} = N_c \gamma$, where $N_c$ is the number of the
pairs of coupling spins, i.e. the number of $\gamma_{jk}$ which are not zero.
Since $\|\wt{A}(t, \omega)\| = \| A(t,\omega) \| / \lambda(t,\omega) = \alpha /
\lambda(t,\omega)$, we need a lower bound of $\lambda(t,\omega)$ to get a
finite $\wt{\alpha}$ as defined in \eqref{eq:Lam_al}. Assume $\lambda(t,\omega)
\geqslant \ell$, and then the estimation \eqref{eq:estimation} becomes
\begin{equation} \label{eq:variance}
\EE_{\Xi} \braket{\Phi_{\Xi}(t)} \leqslant
  \mathrm{e}^{t \Lambda [1 + (\alpha / \ell)^2]} \braket{\Psi(0)}.
\end{equation}
Using the inequality
\begin{equation}
\Lambda [1 + (\alpha / \ell)^2] \geqslant
  \Lambda [1 + (\alpha / \Lambda)^2] \geqslant 2\alpha,
\end{equation}
we get a minimum value for the right hand side of \eqref{eq:variance}, which
can be achieved when $\lambda(t,\omega) \equiv \ell = \Lambda = \alpha$. Such a
choice also turns \eqref{eq:estimation} and \eqref{eq:variance} into
equalities, and we eventually have
\begin{equation} \label{eq:optimal}
\EE_{\Xi} \braket{\Phi_{\Xi}(t)} = \mathrm{e}^{2\alpha t} \braket{\Psi(0)}.
\end{equation}
In fact, we can also show in this case that the variance is minimized
by choosing $\lambda(t, \omega) \equiv \alpha$. To simplify the
notation, we define
\begin{equation}
w_{\Xi}(t) = \exp \left(
  \int_0^t \int_{\Omega} \lambda(s,\omega)
    \,\mathrm{d}\mu_{\omega} \,\mathrm{d}s
\right) \prod_{m=1}^M \frac{\alpha}{\lambda(t_m,\omega_m)}.
\end{equation}
Then $\braket{\Phi_{\Xi}(t)} = [w_{\Xi}(t)]^2 \braket{\Psi(0)}$. Applying
\eqref{eq:expectation_F} to $w_{\Xi}(t)$ yields $\EE_{\Xi} w_{\Xi}(t) =
\mathrm{e}^{\alpha t}$, and thus we get the following estimation of the lower
bound:
\begin{equation}
\begin{split}
\EE_{\Xi} \braket{\Phi_{\Xi}(t)} &=
  \braket{\Psi(0)} \EE_{\Xi} [w_{\Xi}(t)]^2 \\
& \geqslant \braket{\Psi(0)} [\EE_{\Xi} w_{\Xi}(t)]^2 =
  \mathrm{e}^{2 \alpha t} \braket{\Psi(0)}.
\end{split}
\end{equation}
From \eqref{eq:optimal}, it is clear that $\lambda(t,\omega) = \alpha$ gives
the optimal intensity function.

The above analysis also explains the reason for our choice \eqref{eq:lambda}.
When $\gamma_{jk}(t)$ is either zero or a fixed constant $\gamma$ as stated in
the begining of the previous paragraph, we can naturally remove the tuples
$(i,j,k)$ with $\gamma_{jk}(t) \equiv 0$ from the sample space $\Omega$ defined
in \eqref{eq:Omega}, and then $|\Omega| = N_c$ and \eqref{eq:lambda} becomes
$\lambda(t,\omega) = N_c \gamma = \alpha$, which gives exactly the optimal
intensity function.

\section{Numerical results}

\subsection{Examples with few spins: Validity check} \label{sec:ex1}
To verify the algorithm, we first consider some simple cases with only a few
spins. The magnetic field $(m_1(t), m_2(t))$ is chosen as
\begin{displaymath}
m_1(t) = \cos(\omega t), \qquad m_2(t) = \sin(\omega t)
\end{displaymath}
withe $\omega = -0.5$, and $m_3^{(k)} = 1$ is used for all the spins. Assuming
that each spin interacts only with its adjacent spins, and all the interaction
strengths are equal, we have that
\begin{displaymath}
\gamma_{jk}(t) = \gamma_0 \delta_{1,|j-k|}.
\end{displaymath}
Initially, we assume that all the spins have the same state
$\ket{\uparrow}$:
\begin{equation} \label{eq:init}
\ket{\Psi(0)} = \ket{\uparrow}^{\otimes N},
\end{equation}
and we are concerned about the evolution of the probability for the ``all
spin-down'' state:
\begin{equation}
p(t) = \expval{B}{\Psi(t)}, \qquad
  B = \ket{\Psi_{\mathrm{down}}}\bra{\Psi_{\mathrm{down}}},
\end{equation}
where $\ket{\Psi_{\mathrm{down}}} = \ket{\downarrow}^{\otimes N}$. For small
$N$, we use a deterministic Runge-Kutta solver to provide reference solutions.

\begin{figure}[!ht]
\centering
\begin{tabular}{rr}
\subfloat[$N=1$]{\includegraphics[scale=.35]{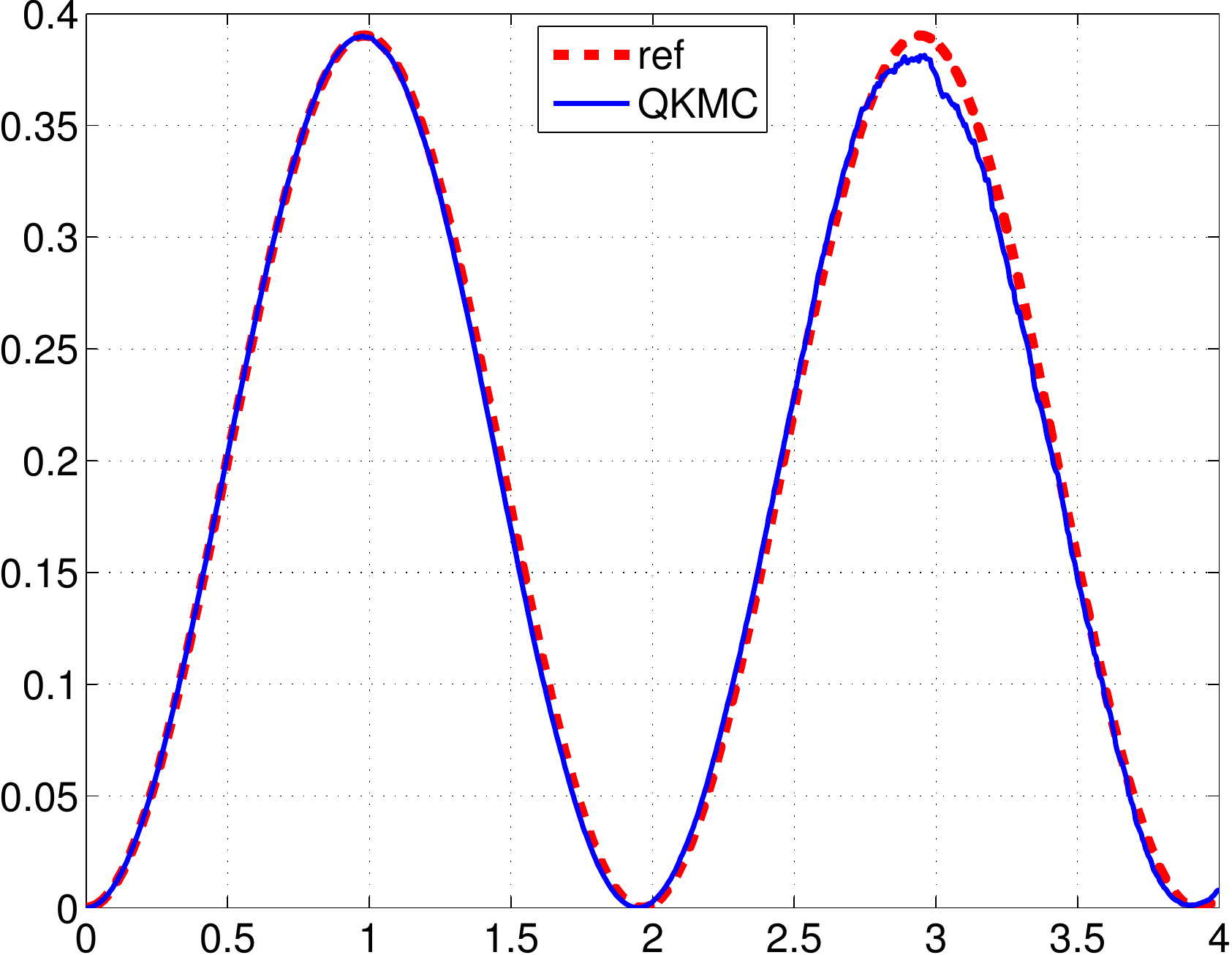}} &
\subfloat[$N=2$]{\includegraphics[scale=.35]{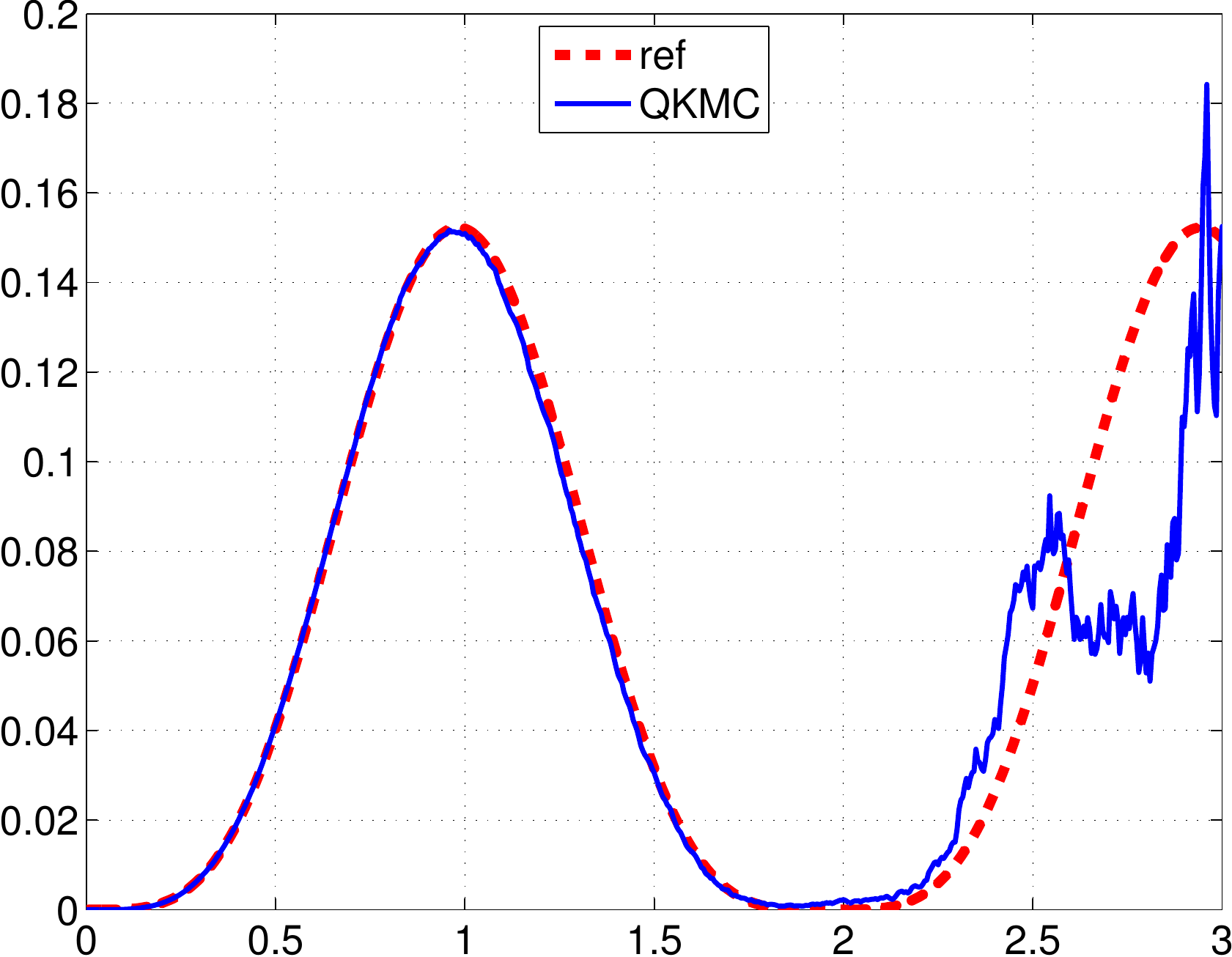}} \\
\subfloat[$N=3$]{\includegraphics[scale=.35]{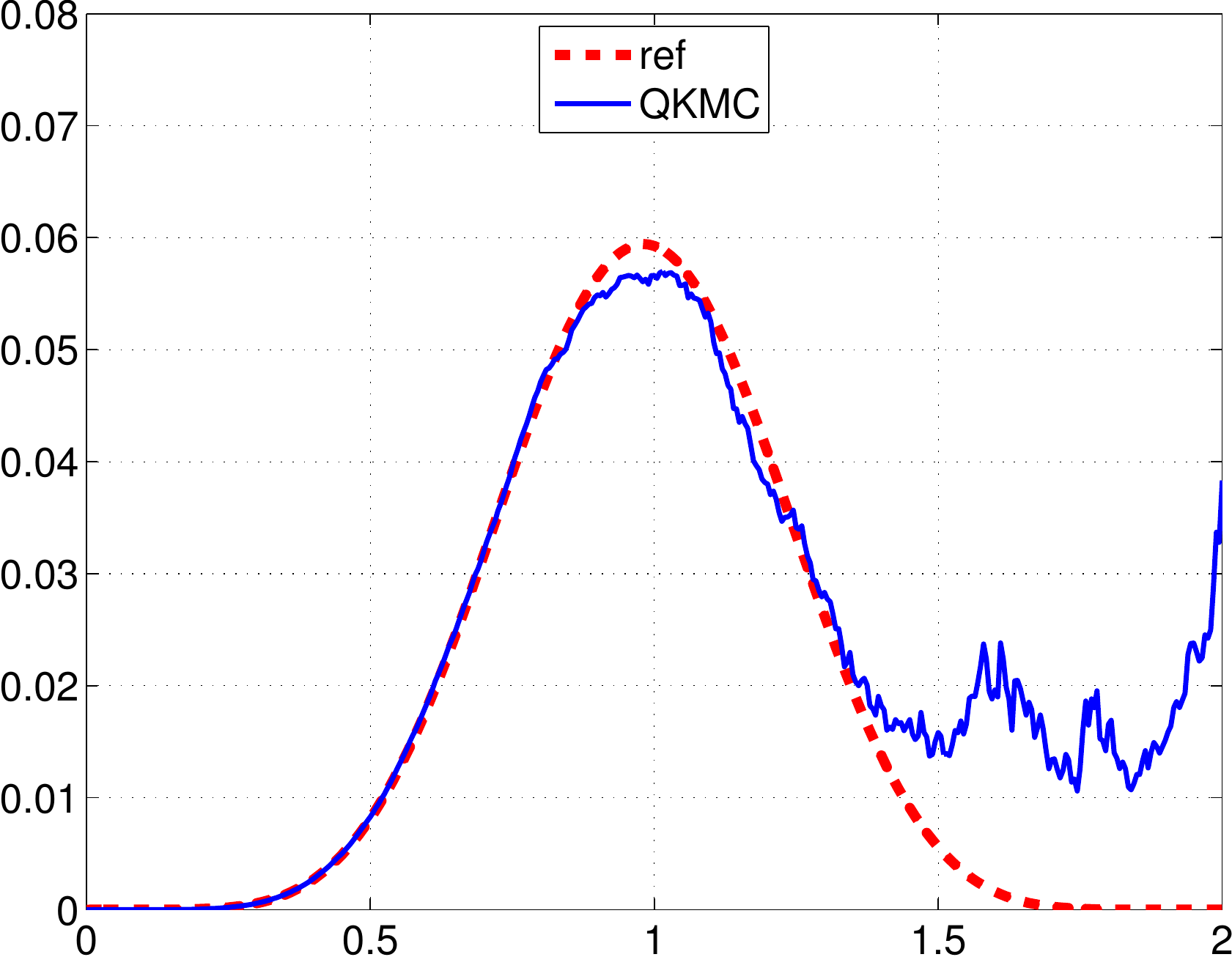}} &
\subfloat[$N=4$]{\includegraphics[scale=.35]{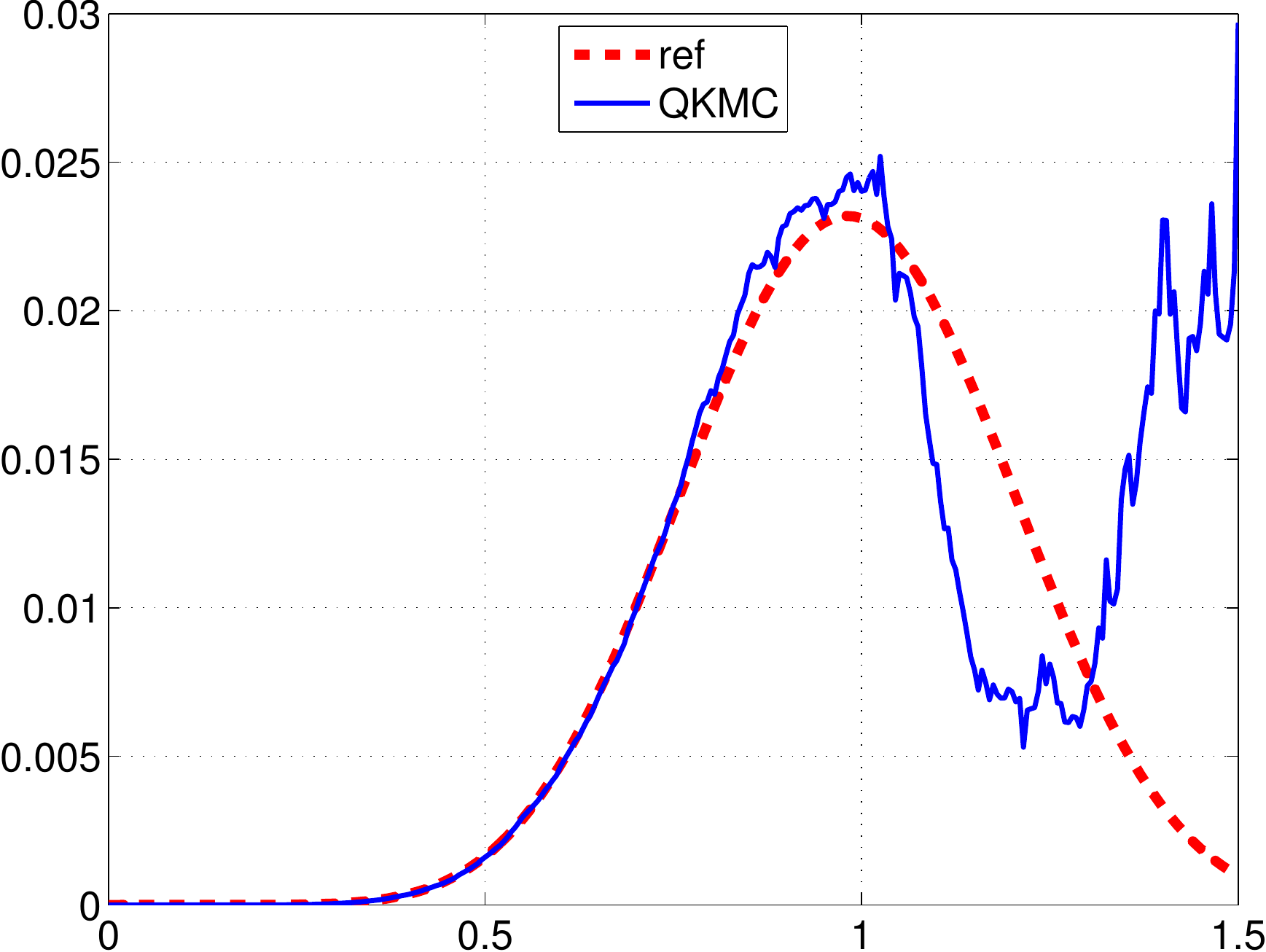}}
\end{tabular}
\caption{Numerical tests of QKMC with $Z$-basis, $\gamma_0 = 0.05$}
\label{fig:1-4spins_Zbasis}
\end{figure}

In Figure \ref{fig:1-4spins_Zbasis}, we show the numerical results using QKMC
with $Z$-basis. The cases with one spin to four spins are considered, and the
coupling intensity $\gamma_0$ is set to be $0.05$. For all the four cases,
1,000,000 trajectories are used. It can be seen that when $t$ is large, the
numerical solution of QKMC becomes oscillatory and unreliable, due to the large
sample variance. As the number of spins increases, the reliable part of the
curve becomes shorter.

\begin{figure}[!ht]
\centering
\begin{tabular}{rr}
\subfloat[$N=1$]{\includegraphics[scale=.35]{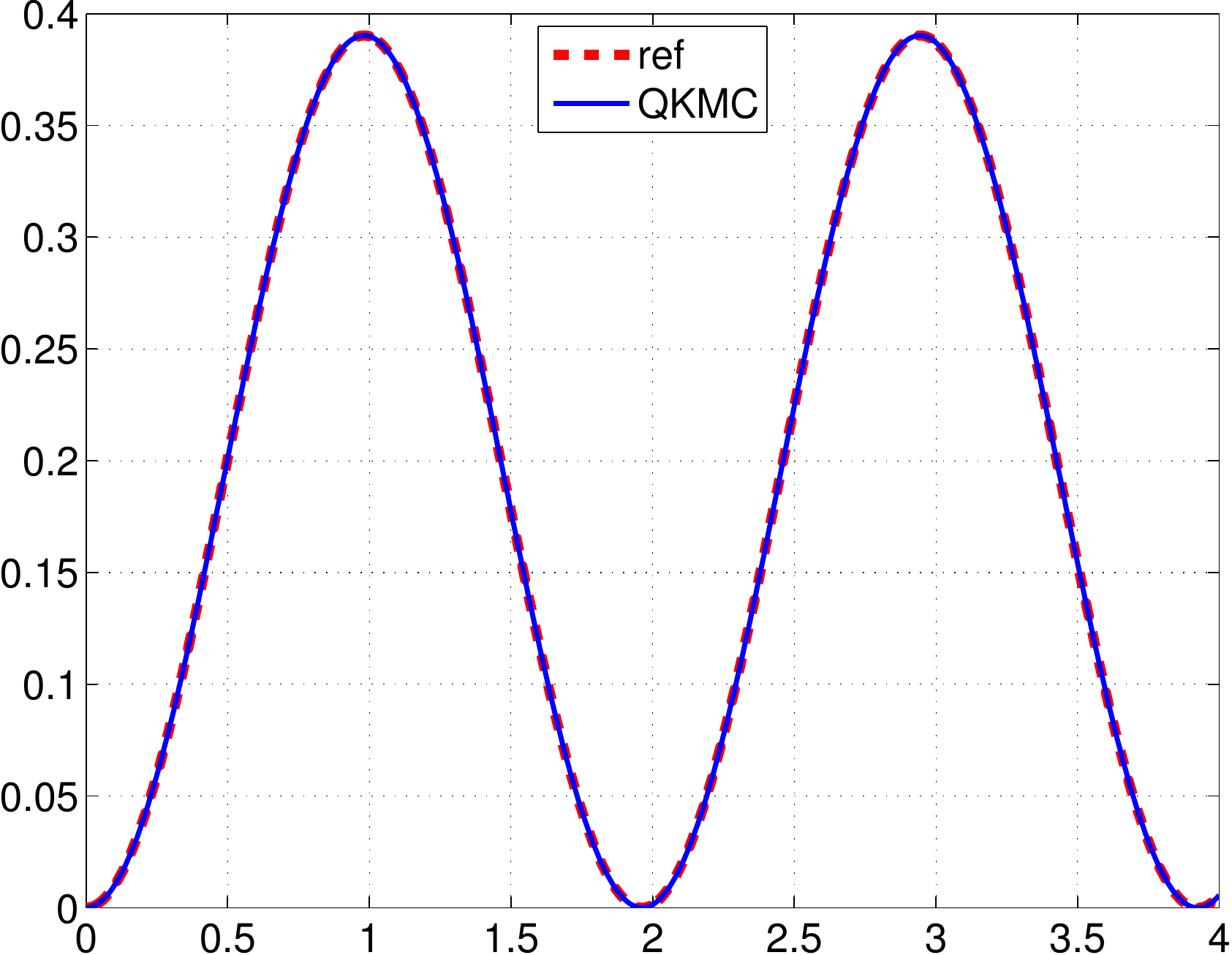}} &
\subfloat[$N=2$]{\includegraphics[scale=.35]{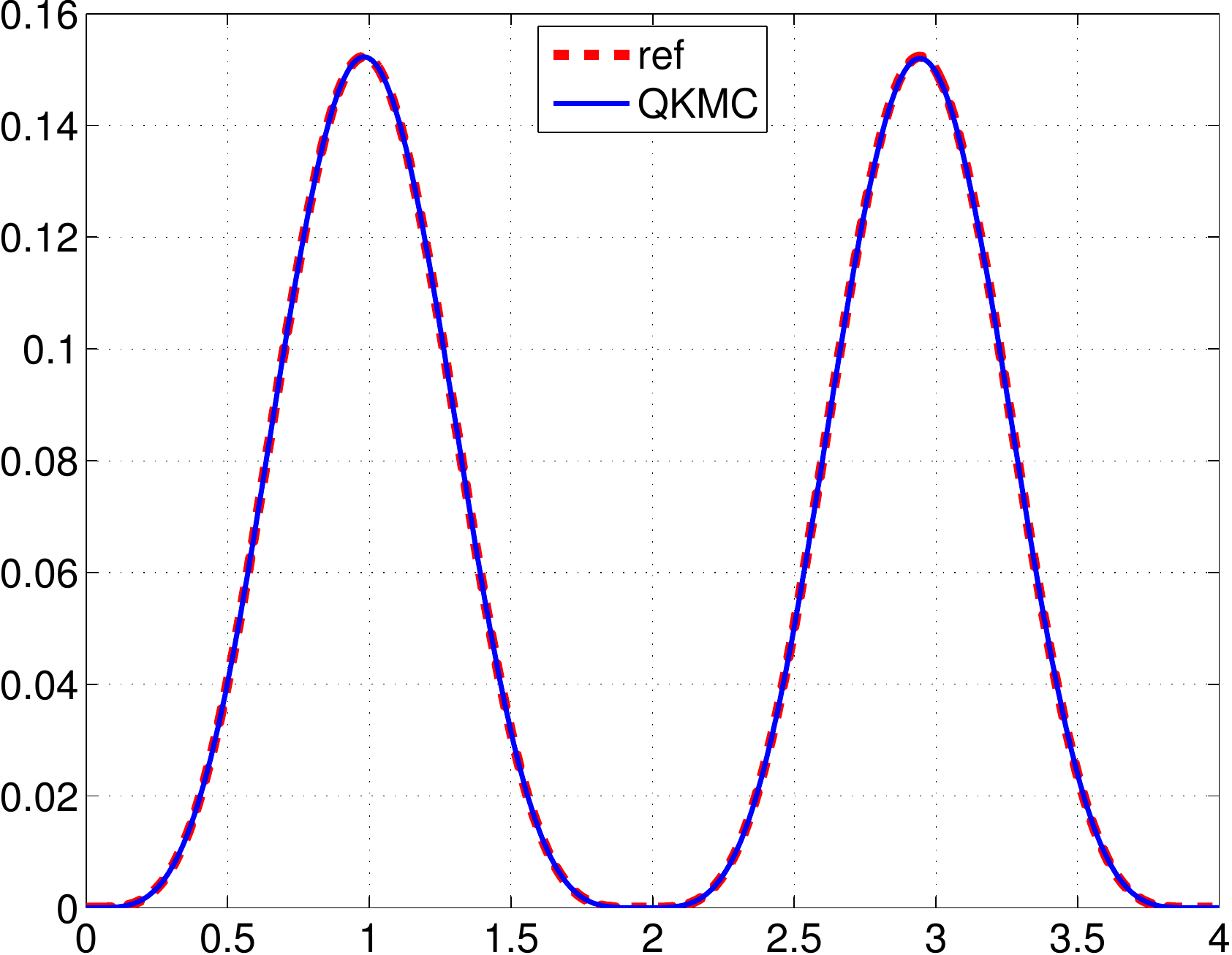}} \\
\subfloat[$N=3$]{\includegraphics[scale=.35]{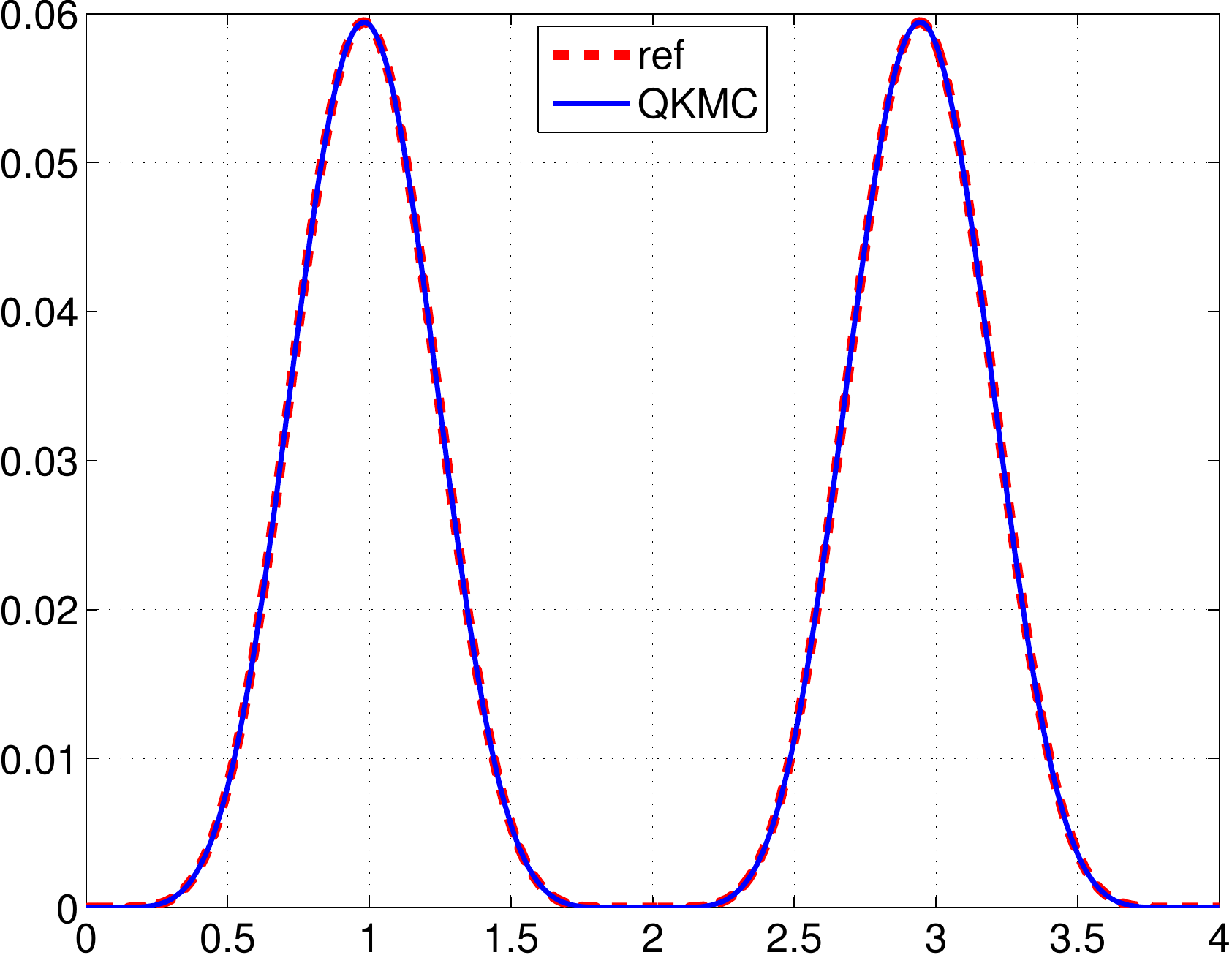}} &
\subfloat[$N=4$]{\includegraphics[scale=.35]{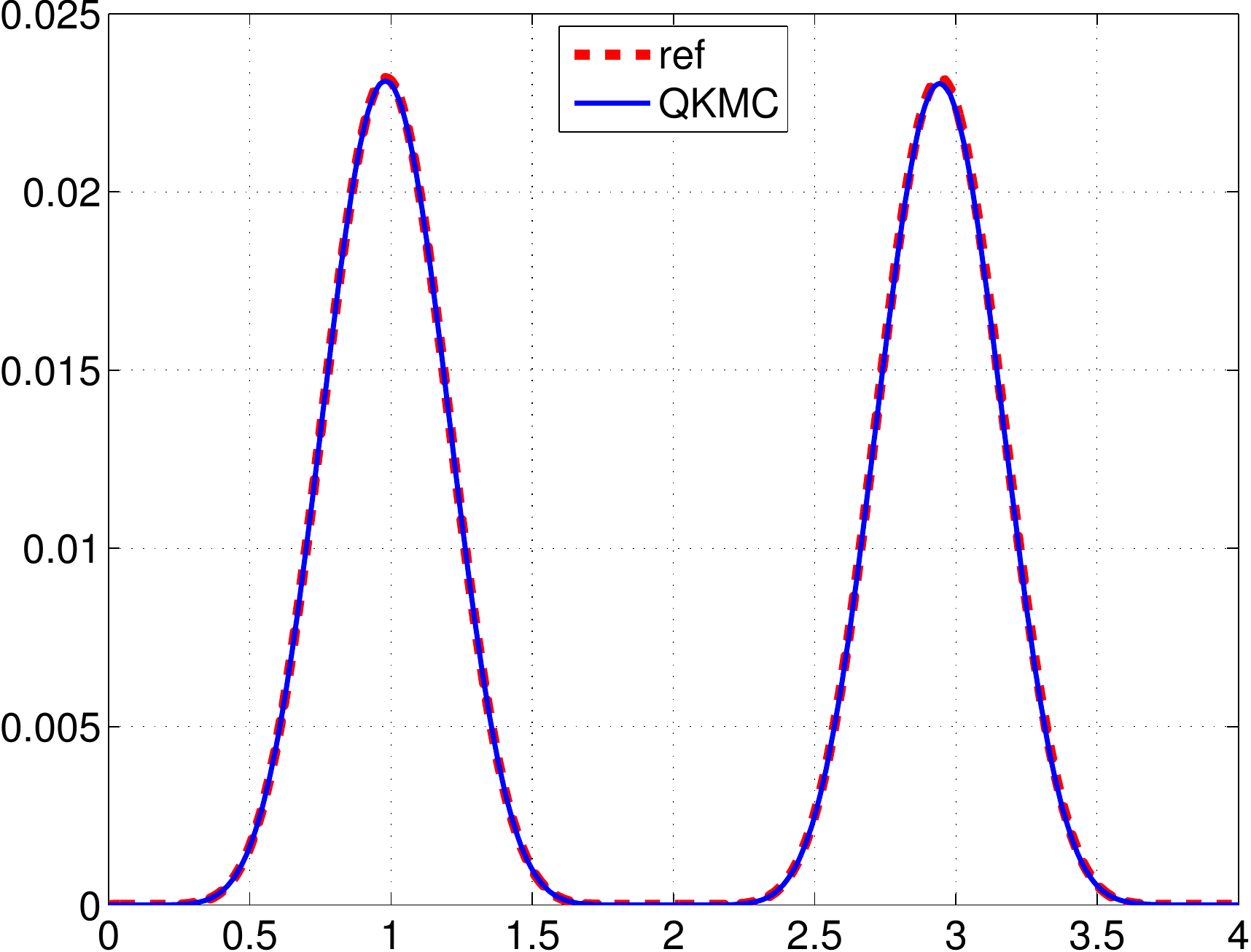}}
\end{tabular}
\caption{Numerical tests of QKMC with simple tensors, $\gamma_0 = 0.05$}
\label{fig:1-4spins}
\end{figure}

Better results can be obtained using the QKMC with simple tensors. Figure
\ref{fig:1-4spins} shows the numerical results with the same settings. The
number of trajectories is again 1,000,000. For this method, when only one spin
is present, the algorithm is identical to the deterministic ODE solver. In all
the four cases, two complete cycles are obtained without obvious oscillation.

\begin{figure}[!ht]
\centering
\includegraphics[width=.6\textwidth]{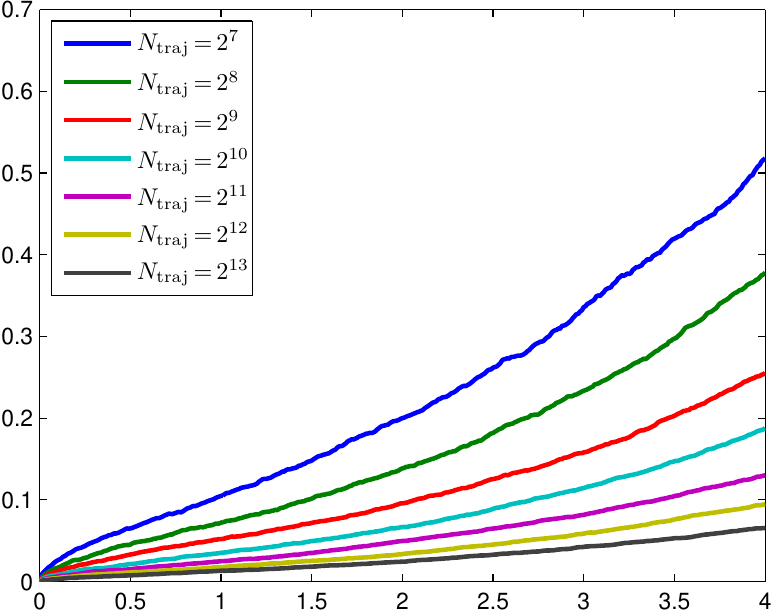}
\caption{Evolution of the numerical error $e_{N_{\mathrm{traj}}}$ (defined in
\eqref{eq:error}) for 4 spins. The $x$-axis is the time $t$, and the $y$-axis
is the 2-norm error of the solution.}
\label{fig:error}
\end{figure}

\begin{figure}[!ht]
\centering
\includegraphics[width=.6\textwidth]{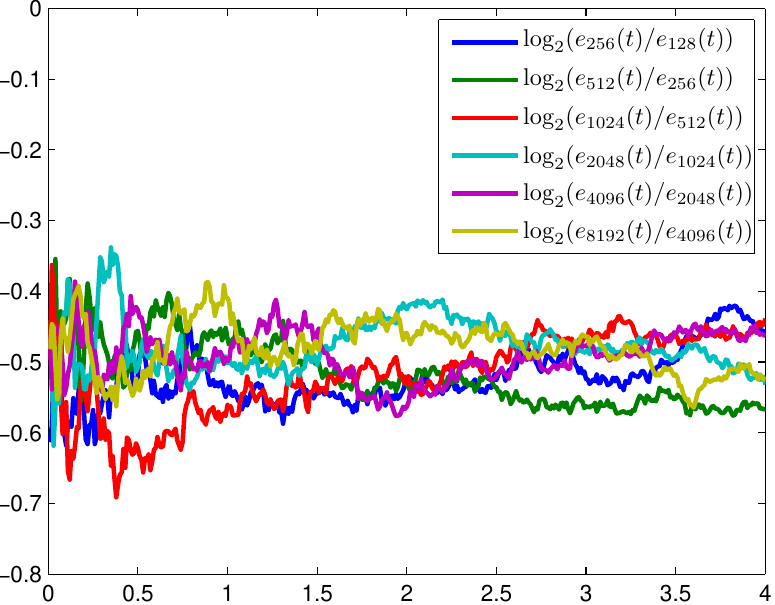}
\caption{Evolution of the numerical order for 4 spins. The $x$-axis is the time
$t$, and the $y$-axis is the numerical order obtained as indicated in the
legend.}
\label{fig:order}
\end{figure}

In the case of 4 spins, we also check the numerical error for QKMC with simple
tensors. Using the result of the deterministic solver as the reference solution
$\ket{\Psi^{\mathrm{ref}}(t)}$, we plot the evolution of the 2-norm of
numerical error $\ket{\Psi^{\mathrm{err}}(t)} = \ket{\Psi^{\mathrm{num}}(t)} -
\ket{\Psi^{\mathrm{ref}}(t)}$:
\begin{equation} \label{eq:error}
e_{N_{\mathrm{traj}}}(t) = \sqrt{\braket{\Psi^{\mathrm{err}}(t)}},
\end{equation}
where $\ket{\Psi^{\mathrm{num}}(t)}$ is the numerical solution of QKMC with
$N_{\mathrm{traj}}$ trajectories. Figure \ref{fig:error} shows the reduction of
the error as the number of trajectories increases, and the order of convergence
is calculated in Figure \ref{fig:order}. It is obvious that the numerical order
is around $1/2$ for all $t$, which indicates that the expected convergence rate
in the Monte Carlo method is achieved in our numerical test.

\subsection{High dimensional tests}

\begin{figure}[!ht]
\centering
\begin{tabular}{rr}
\subfloat[$N=10$]{\includegraphics[scale=.36]{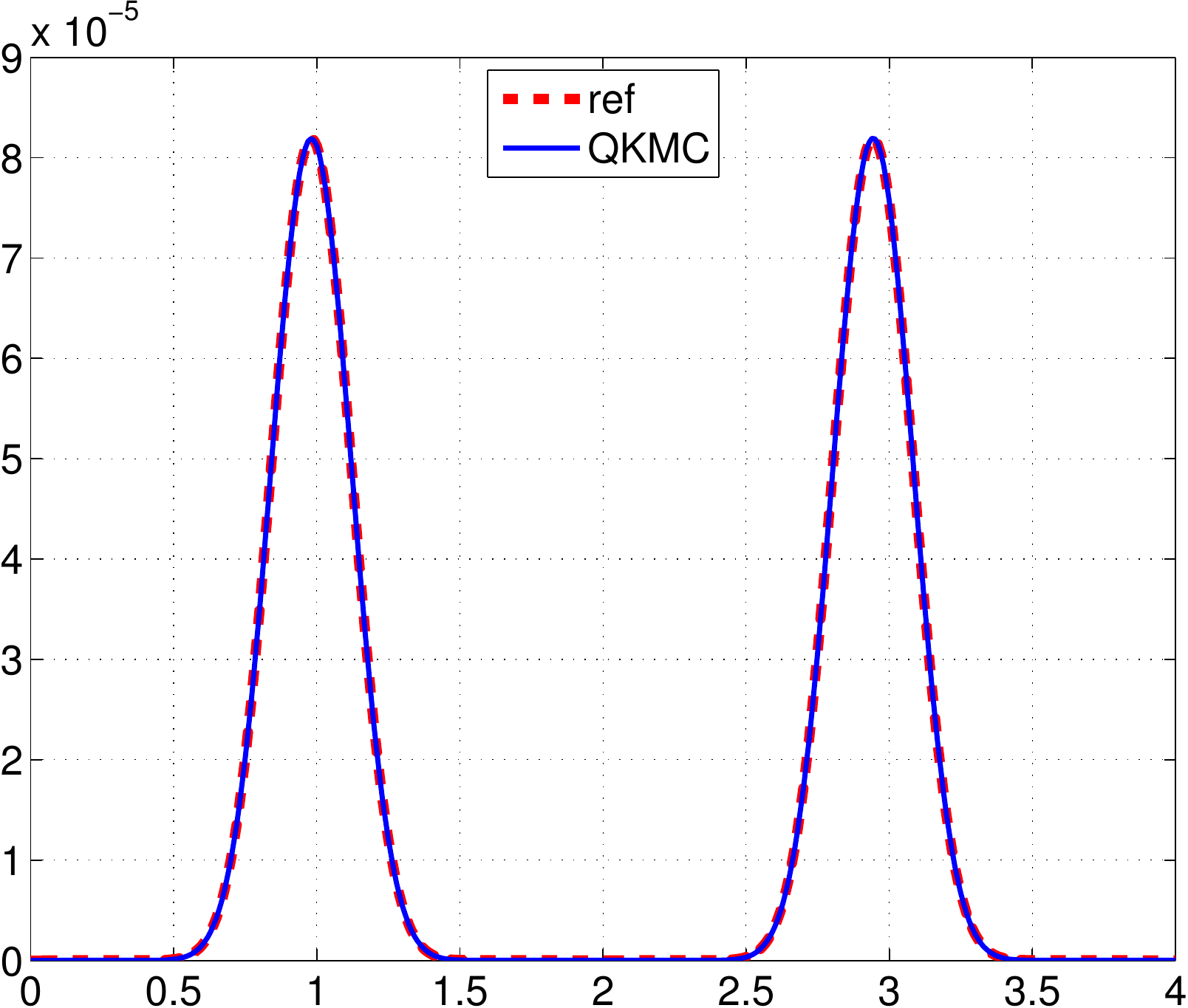}} &
\subfloat[$N=20$]{\includegraphics[scale=.36]{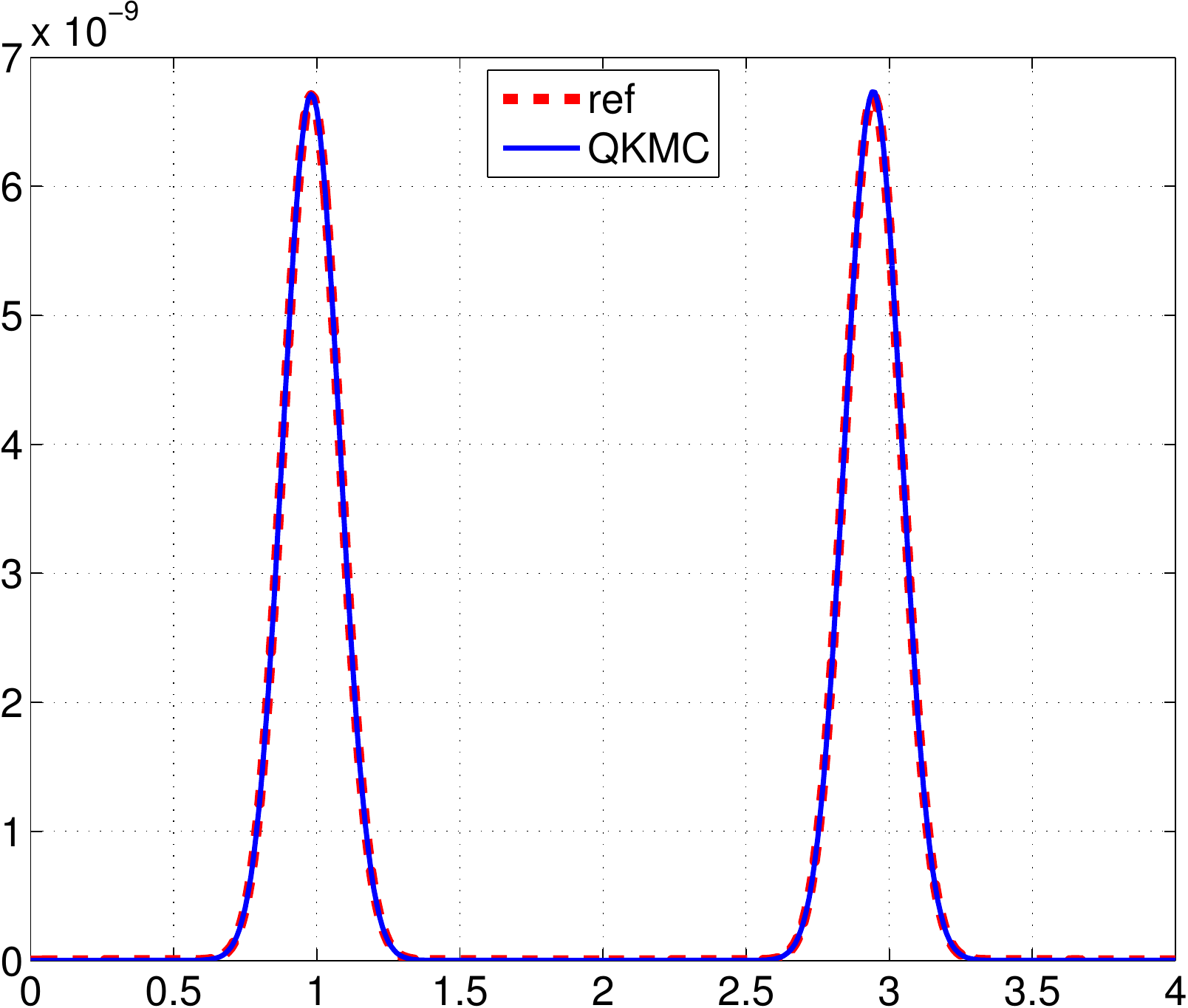}} \\
\subfloat[$N=30$]{\includegraphics[scale=.36]{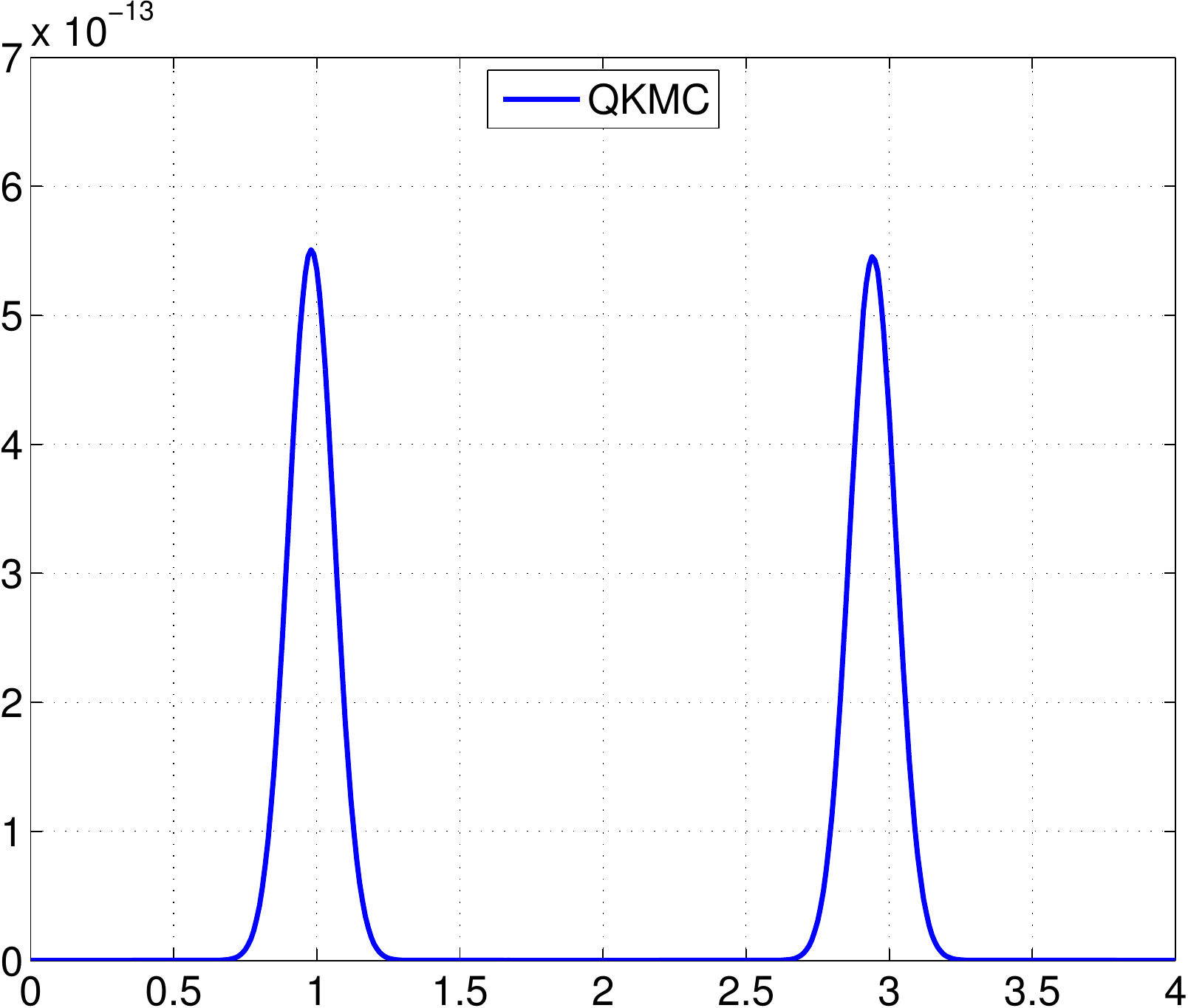}} &
\subfloat[$N=40$]{\includegraphics[scale=.36]{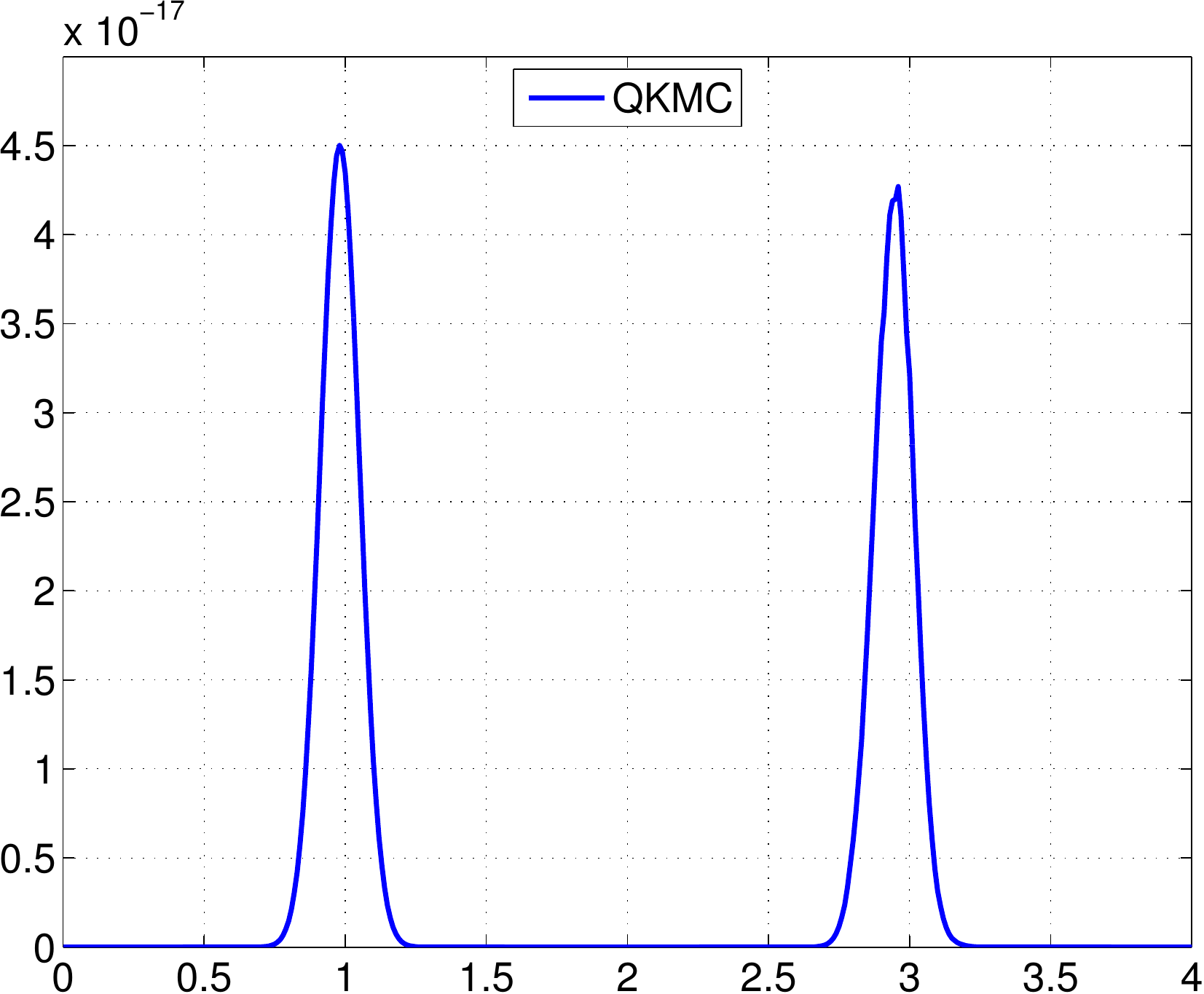}}
\end{tabular}
\caption{Numerical tests of QKMC with simple tensors, $\gamma_0 = 0.01$}
\label{fig:10-40spins}
\end{figure}

The cases with more spins are shown in Figure
\ref{fig:10-40spins}. Here we consider weaker coupling intensity
$\gamma_0 = 0.01$, while the number of spins ranges from 10 to 40. In
these cases, the method with $Z$-basis needs a huge number of
trajectories to get meaningful results, which is out of our current
computational capacity. However, by using $8,000,000$ trajectories,
two complete cycles are still well obtained using QKMC with simple
tensors. For 10 and 20 spins, the numerical results are again
validated by comparison with the reference results. For 30 and 40
spins, the reference solutions are not provided since the
computational time for a deterministic solver is not affordable. For
the case of 40 spins, the second peak is lower than the first one,
which might indicate some numerical error induced by insufficient
number of trajectories.

\subsection{Numerical examples with random parameters}
In this example, we assume that the interactions and the energy splittings are
random. For spin interactions, we let $\mc{I}_c$ be the set of all possible
interaction pairs:
\begin{equation}
\mc{I}_c = \{(j,k) \mid 1 \le j < k \le N\}.
\end{equation}
For each trajectory, we randomly pick a subset $\mc{I} \subset \mc{I}_c$ with
$N-1$ elements. Using $U(a,b)$ to denote the uniform distribution in $[a,b]$,
we set the coupling intensity to be
\begin{equation}
\gamma_{jk}(t) \left\{ \begin{array}{ll}
  \sim U(\gamma_0 - \Delta\gamma, \gamma_0 + \Delta\gamma), &
    \text{if } (j,k) \in \mc{I}, \\
  = 0, & \text{otherwise.}
\end{array} \right.
\end{equation}
The energy splitting $m_3^{(k)}$ is also assumed to be uniformly distributed:
\begin{equation}
m_3^{(k)} \sim U(0.9, 1.1).
\end{equation}
The initial condition is the same as \eqref{eq:init}.

\begin{figure}[!ht]
\centering
\begin{tabular}{rr}
\subfloat[$N=1$]{\includegraphics[scale=.35]{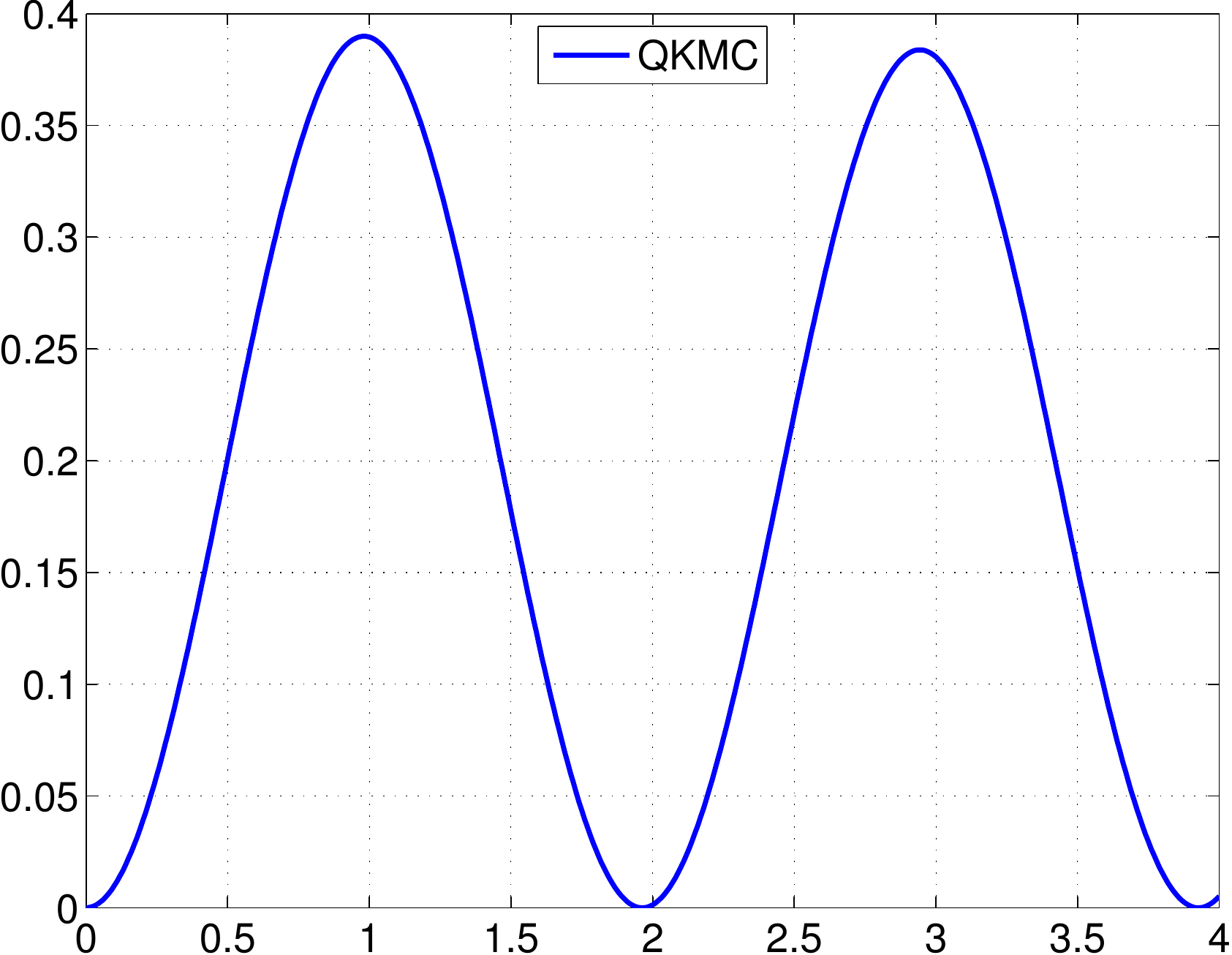}} &
\subfloat[$N=2$]{\includegraphics[scale=.35]{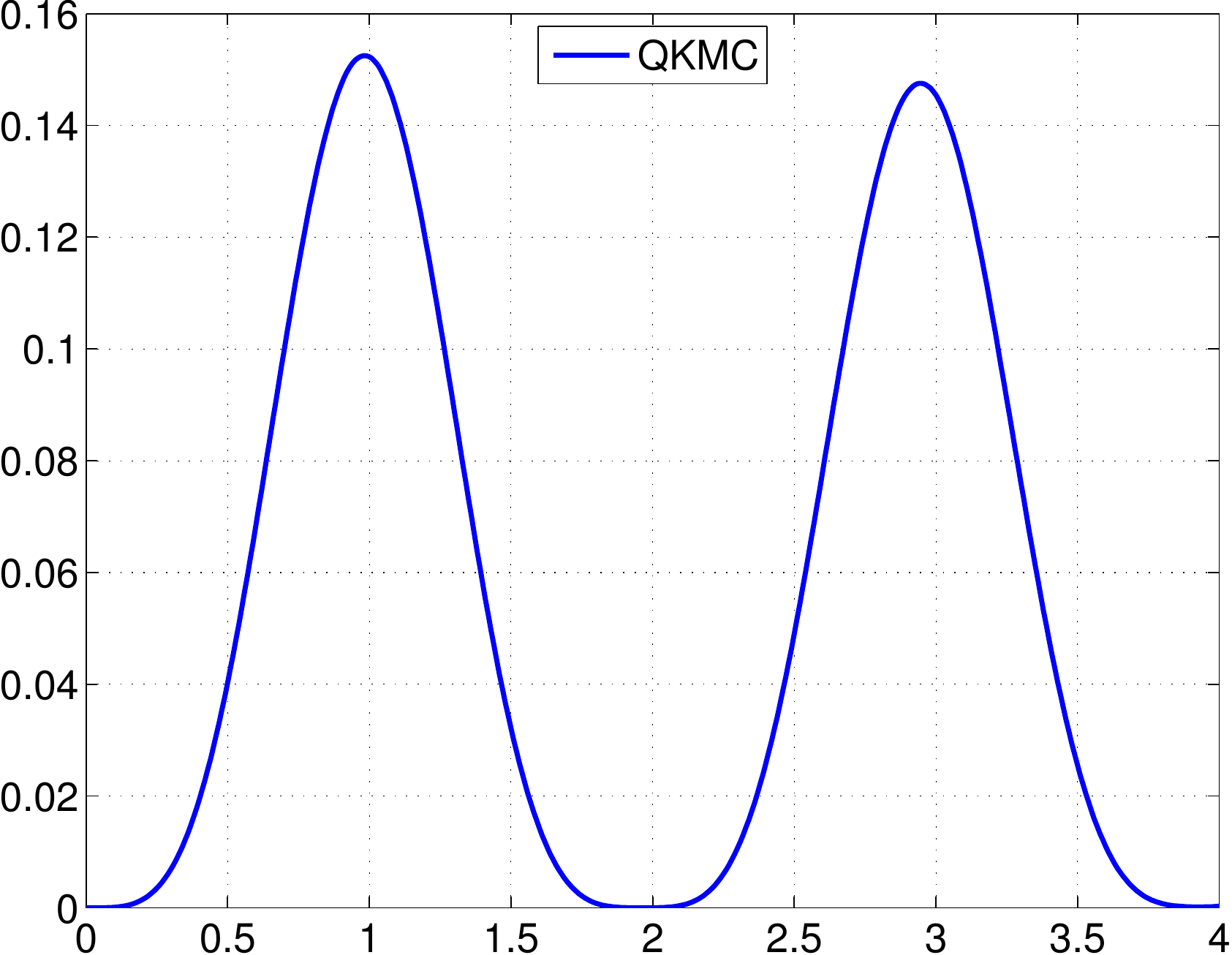}} \\
\subfloat[$N=3$]{\includegraphics[scale=.35]{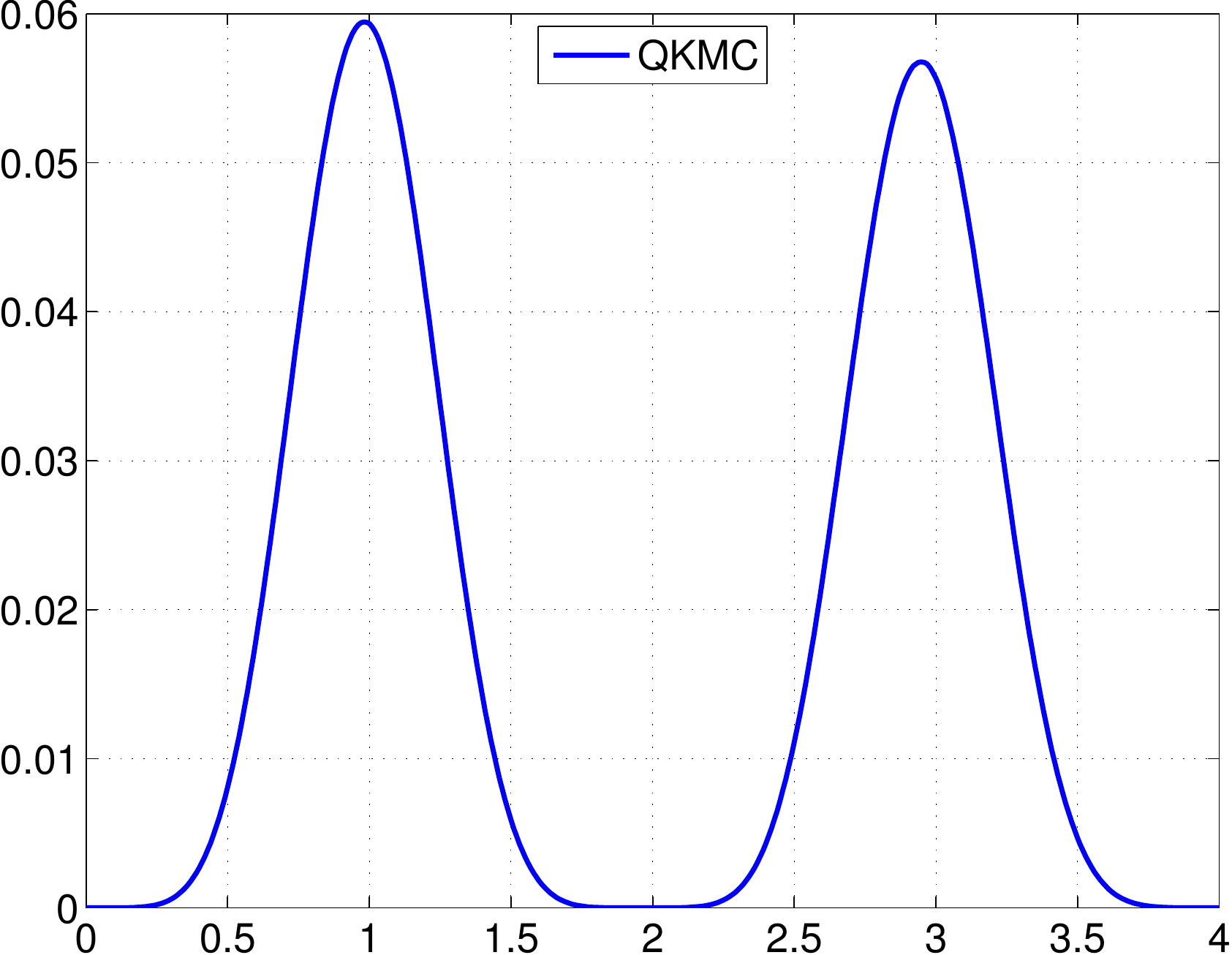}} &
\subfloat[$N=4$]{\includegraphics[scale=.35]{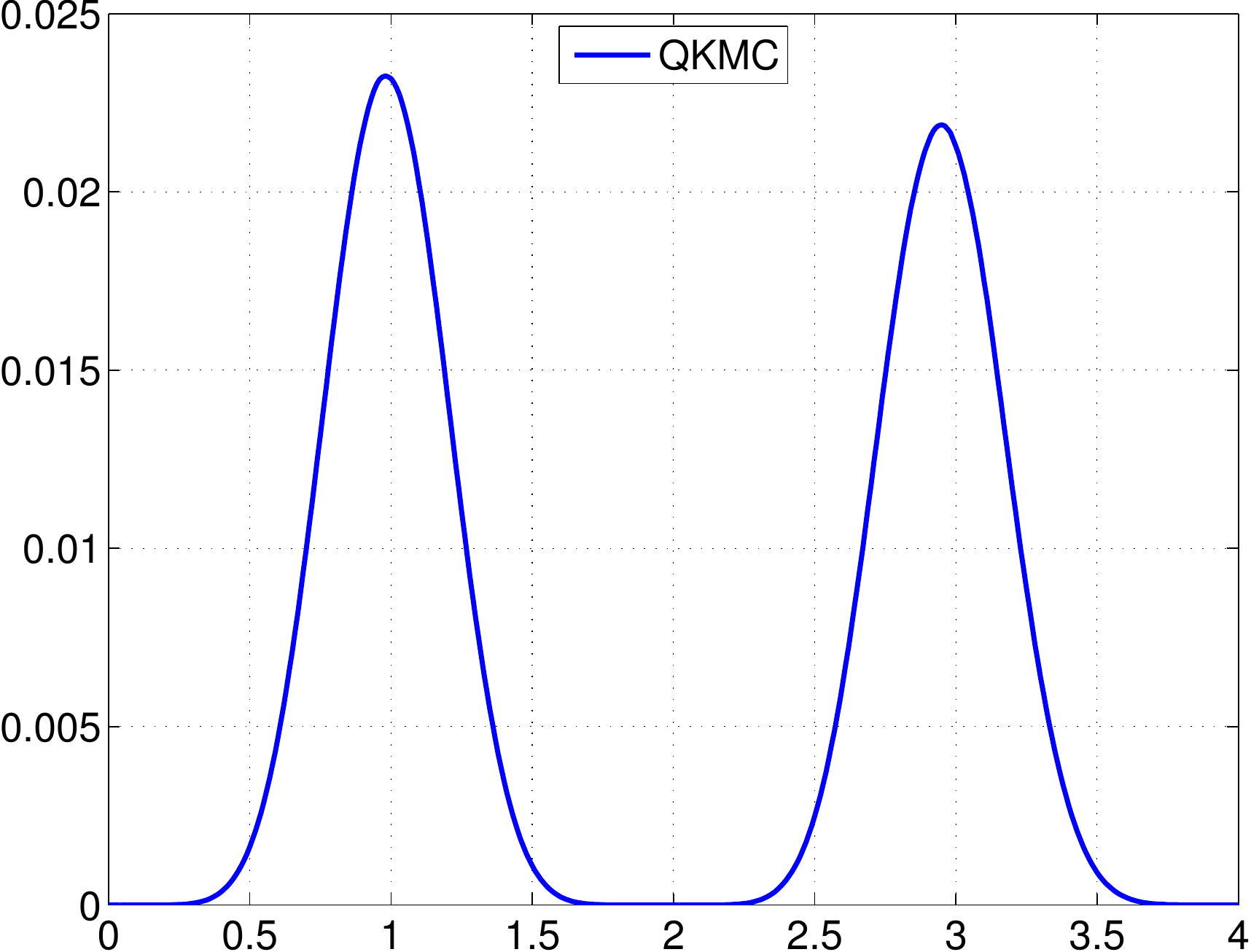}}
\end{tabular}
\caption{Numerical tests with randomness, $\gamma_0 = 0.05$}
\label{fig:1-4spins_random}
\end{figure}

\begin{figure}[!ht]
\centering
\begin{tabular}{rr}
\subfloat[$N=10$]{\includegraphics[scale=.36]{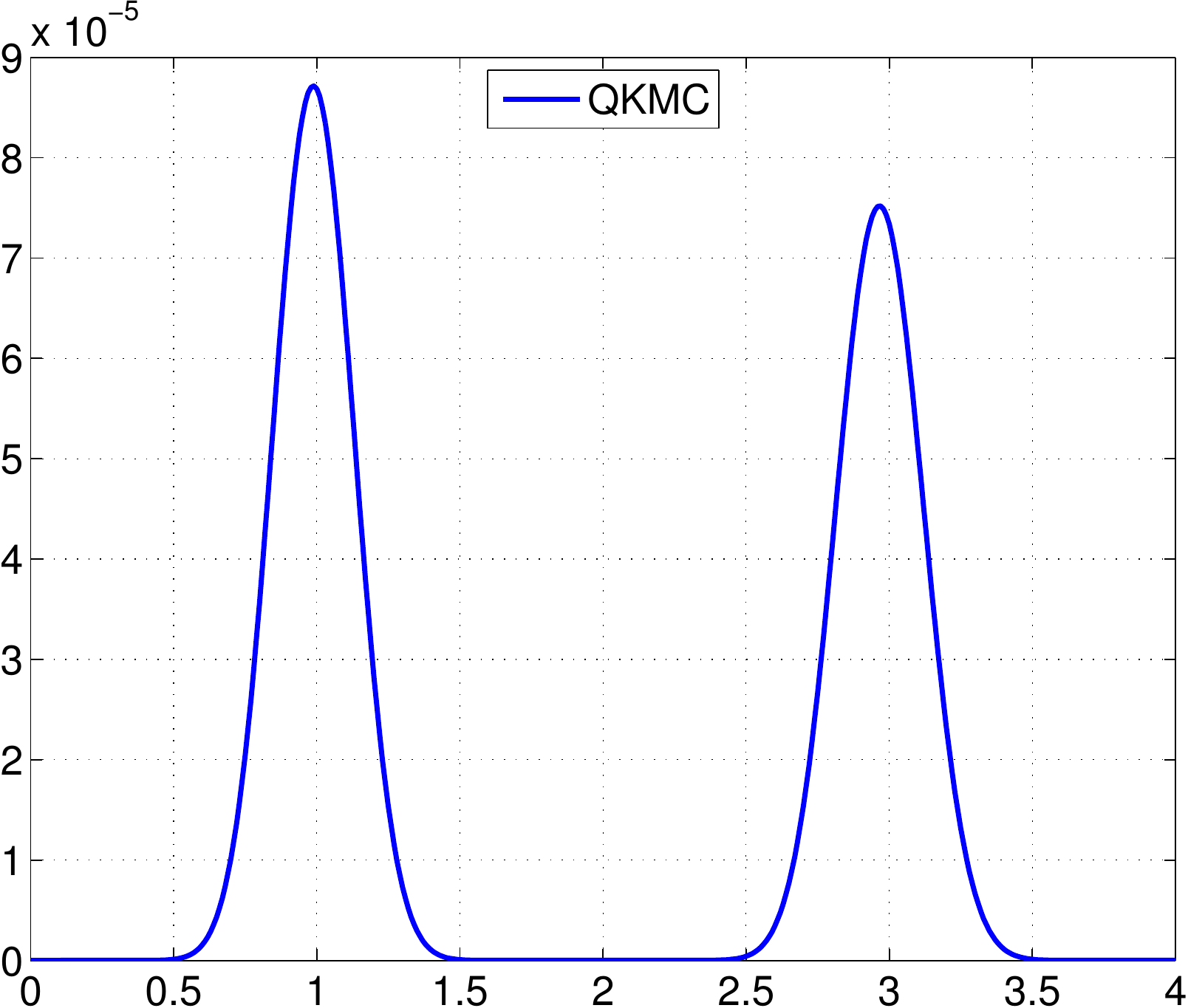}} &
\subfloat[$N=20$]{\includegraphics[scale=.36]{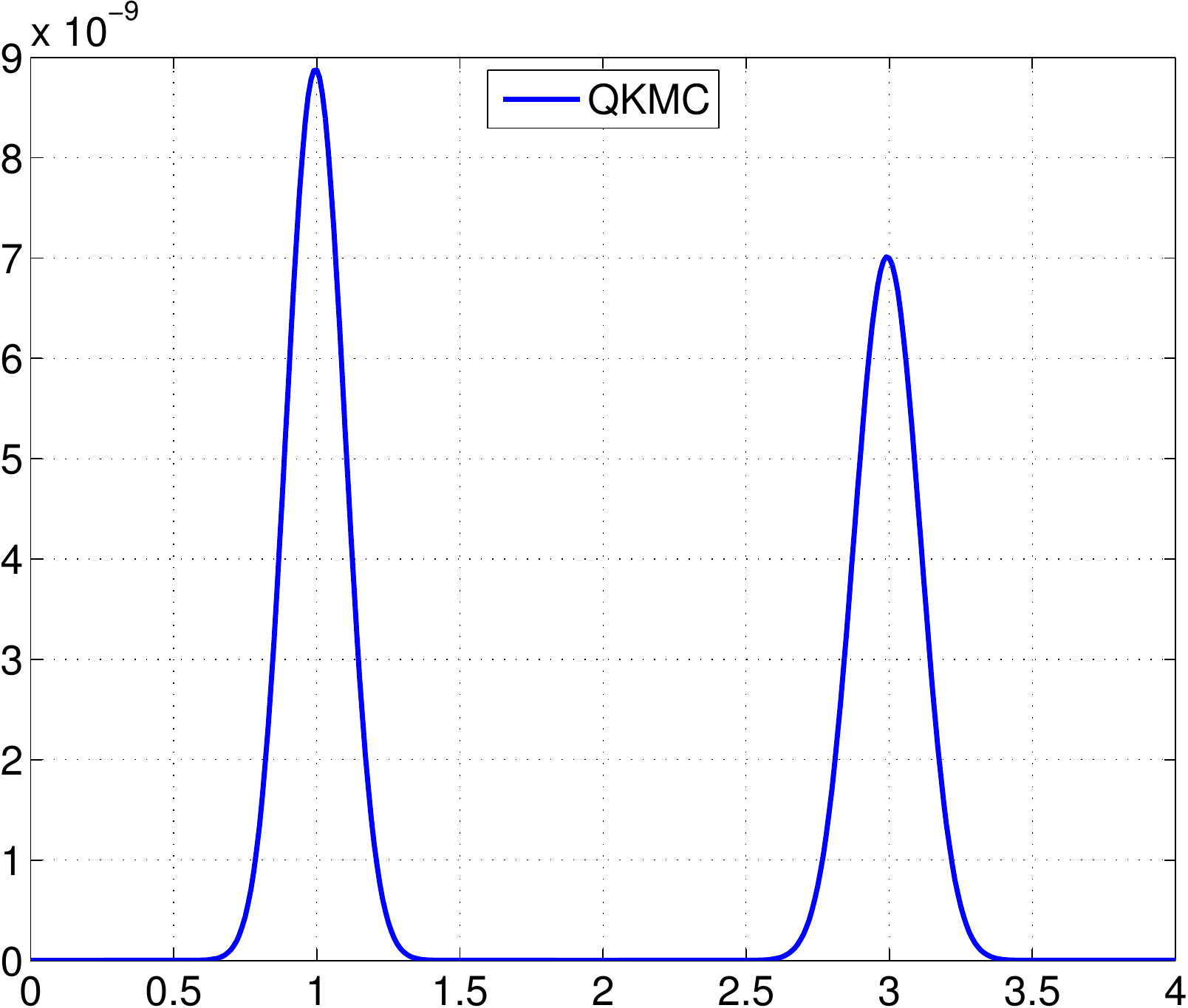}} \\
\subfloat[$N=30$]{\includegraphics[scale=.36]{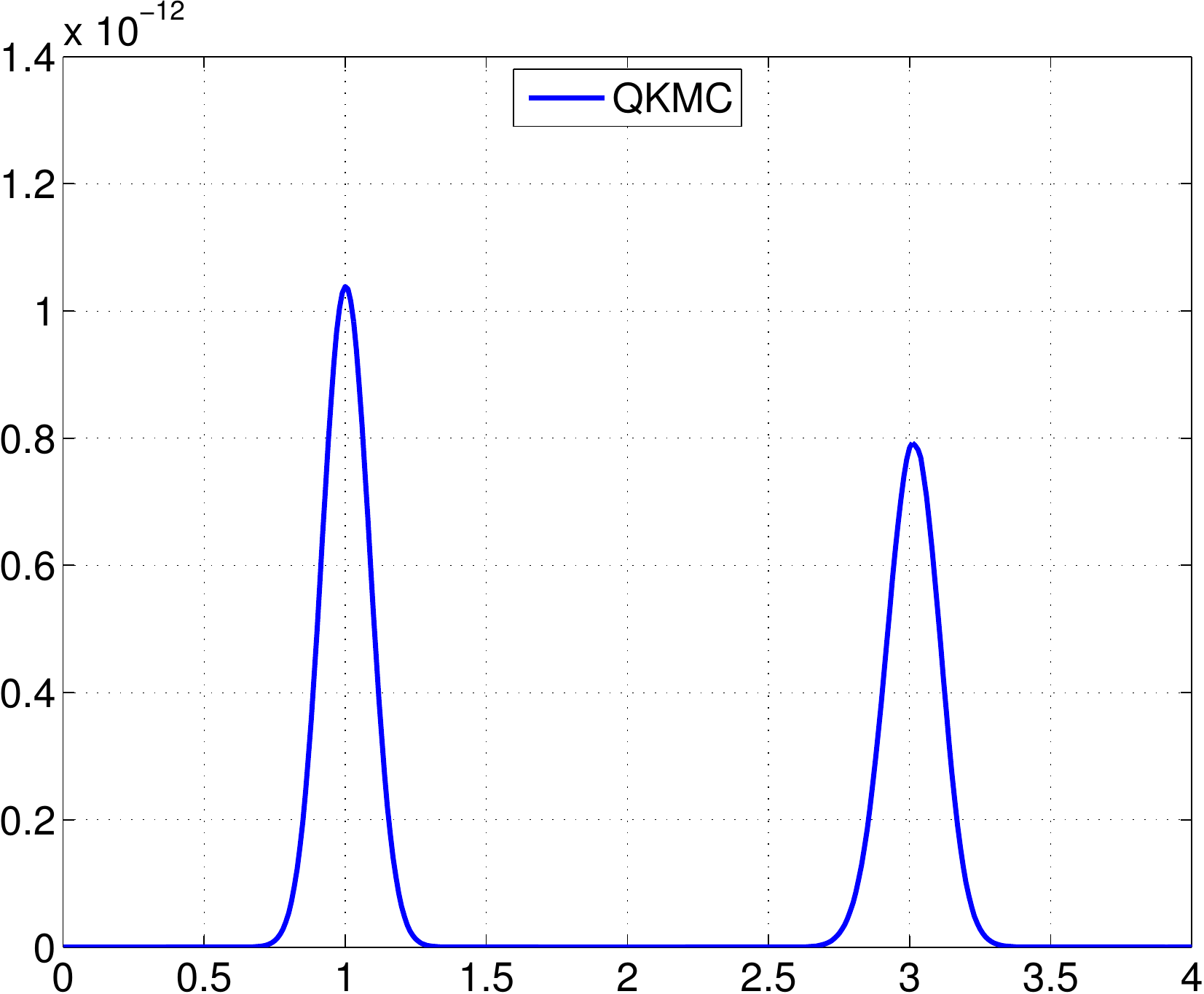}} &
\subfloat[$N=40$]{\includegraphics[scale=.36]{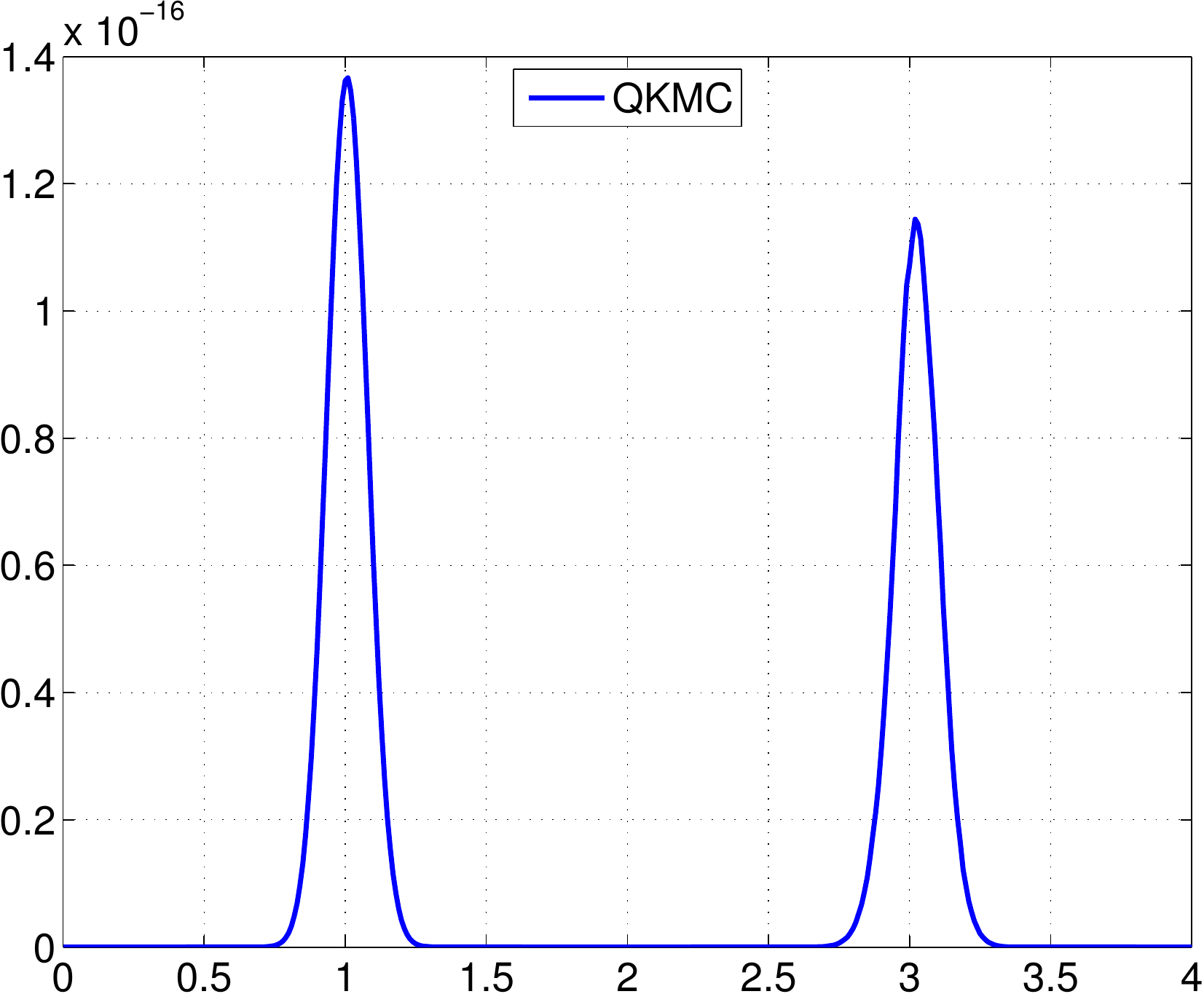}}
\end{tabular}
\caption{Numerical tests with randomness, $\gamma_0 = 0.01$}
\label{fig:10-40spins_random}
\end{figure}

Figure \ref{fig:1-4spins_random} shows the results of $p(t)$ with $\gamma_0 =
0.05$ and $\Delta\gamma = 0.005$ for one to four spins, and the number of
trajectories is again $1,000,000$. Different from the deterministic cases in
Section \ref{sec:ex1}, the second peak is slightly lower than the first peak,
which indicates some cancellation between different interaction patterns.
Similarly, we also consider the case with weaker coupling intensity $\gamma_0 =
0.01$, $\Delta\gamma = 0.001$ but more spins. Smooth results can be obtained
with $8,000,000$ trajectories for as many as $40$ spins. Figure
\ref{fig:10-40spins_random} also shows the lower second peaks, which indicates
the qualitatively correct behavior of the numerical solution.

\section{Conclusion}
In this work, we propose a stochastic method to solve the multi-spin
dynamics.  The method is derived from a quite general framework and
the multi-spin dynamics appears here to be an interesting
application. Numerical experiments show that very small ``flipping
probabilites'' can be well captured by this method in a system with as
many as 40 spins. Future work includes other applications of this
framework and numerical techniques reducing the variance in the
solution and better preserving the conservative quantities.

\bibliographystyle{siamplain}
\bibliography{SpinDynamics}

\begin{thebibliography}{10}

\bibitem{CaiLu2017}
{\sc Z.~Cai and J.~Lu}, {\em A surface hopping {G}aussian beam method for
  high-dimensional transport systems}, 2017.
\newblock preprint, arXiv:1703.06116.

\bibitem{CohenGullReichmanMillis:2015}
{\sc G.~Cohen, E.~Gull, D.~R. Reichman, and A.~J. Millis}, {\em Taming the
  dynamical sign problem in real-time evolution of quantum many-body problems},
  Phys. Rev. Lett., 115 (2015), p.~266802.

\bibitem{Ernst:90}
{\sc R.~R. Ernst, G.~Bodenhausen, and A.~Wokaun}, {\em Principles of Nuclear
  Magnetic Resonance in One and Two Dimensions}, Claredon Press, 1990.

\bibitem{FrieseckeHennekeKunisch}
{\sc G.~Friesecke, F.~Henneke, and K.~Kunisch}, {\em Sparse control of quantum
  systems}, 2015.
\newblock preprint, arXiv:1507.00768.

\bibitem{Gull:2011}
{\sc E.~Gull, A.~Millis, A.~Lichtenstein, A.~Rubtsov, M.~Troyer, and
  P.~Werner}, {\em Continuous-time {M}onte {C}arlo methods for quantum impurity
  models}, Rev. Mod. Phys., 83 (2011), pp.~349--404.

\bibitem{HirschFye:1986}
{\sc J.~E. Hirsch and R.~M. Fye}, {\em {M}onte {C}arlo method for magnetic
  impurities in metals}, Phys. Rev. Lett., 56 (1986), p.~2521.

\bibitem{Hubbard:1959}
{\sc J.~Hubbard}, {\em Calculation of partition functions}, Phys. Rev. Lett., 3
  (1959), p.~77.

\bibitem{Jacobsen2006}
{\sc M.~Jacobsen}, {\em Point Process Theory and Applications}, Probability and
  Its Applications, Birkh{\"a}user Basel, 2006.

\bibitem{LuZhou2016}
{\sc J.~Lu and Z.~Zhou}, {\em Frozen {G}aussian approximation with surface
  hopping for mixed quantum-classical dynamics: {A} mathematical justification
  of fewest switches surface hopping algorithms}, Math. Comp., in press.
\newblock arXiv:1602.06459.

\bibitem{MeyerMantheCederbaum:1990}
{\sc H.~D. Meyer, U.~Manthe, and L.~S. Cederbaum}, {\em The
  multi-configurational time-dependent {H}artree approach}, Chem. Phys. Lett.,
  165 (1990), p.~73.

\bibitem{MuhlbacherRabani:2008}
{\sc L.~Muhlbacher and E.~Rabani}, {\em Real-time path integral approach to
  nonequilibrium many-body quantum systems}, Phys. Rev. Lett., 100 (2008),
  p.~176403.

\bibitem{Prokofev:96}
{\sc N.~V. Prokof'ev, B.~V. Svistunov, and I.~S. Tupitsyn}, {\em Exact quantum
  {M}onte {C}arlo process for the statistics of discrete systems}, JETP Lett.,
  64 (1996), p.~911.

\bibitem{RubtsovSavkinLichtenstein:2005}
{\sc A.~N. Rubtsov, V.~V. Savkin, and A.~I. Lichtenstein}, {\em Continuous-time
  quantum {M}onte {C}arlo method for fermions}, Phys. Rev. B, 72 (2005),
  p.~035122.

\bibitem{Schachenmayer:2015}
{\sc J.~Schachenmayer, A.~Pikovski, and A.~M. Rey}, {\em Many-body quantum spin
  dynamics with monte carlo trajectories on a discrete phase space}, Phys. Rev.
  X, 5 (2015), p.~011022.

\bibitem{Schollwock:2011}
{\sc U.~Schollw\"ock}, {\em The density-matrix renormalization group in the age
  of matrix product states}, Ann. Phys., 326 (2011), pp.~96--192.

\bibitem{Stratonovich:1958}
{\sc R.~L. Stratonovich}, {\em On a method of calculating quantum distribution
  functions}, Soviet Physics Doklady, 2 (1958), p.~416.

\bibitem{Vidal:2003}
{\sc G.~Vidal}, {\em Efficient classical simulation of slightly entangled
  quantum computations}, Phys. Rev. Lett., 91 (2003), p.~147902.

\bibitem{White:1992}
{\sc S.~R. White}, {\em Density matrix formulation for quantum renormalization
  groups}, Phys. Rev. Lett., 69 (1992), p.~2863.

\bibitem{Zhang:2013}
{\sc S.~Zhang}, {\em Auxiliary-field quantum {M}onte {C}arlo for correlated
  electron systems}, in {Emergent Phenomena in Correlated Matter, Modeling and
  Simulation Vol. 3}, Forschungszentrum J\"ulich, 2013.

\bibitem{ZhangKrakauer:2003}
{\sc S.~Zhang and H.~Krakauer}, {\em Quantum {M}onte {C}arlo method using
  phase-free random walks with {S}later determinants}, Phys. Rev. Lett., 90
  (2003), p.~136401.

\end{thebibliography}

\end{document}